\documentclass[a4paper,11pt]{article}
\usepackage{graphicx,rotating,hyperref,slashed,amsmath,xcolor,amssymb,amsfonts,colortbl,cite}
\pdfoutput=1
\makeatletter
\hypersetup{colorlinks,bookmarksopen,bookmarksnumbered,
linkcolor=blus,pdfstartview=FitH,urlcolor=rossos,citecolor=verde}
\allowdisplaybreaks

\def\lsim{\mathrel{\rlap{\lower3pt\hbox{\hskip0pt$\sim$}}
   \raise1pt\hbox{$<$}}}         
\def\gsim{\mathrel{\rlap{\lower4pt\hbox{\hskip1pt$\sim$}}
   \raise1pt\hbox{$>$}}}         

 \newcommand{\sfootnote}[1]{}
\definecolor{bluc}{cmyk}{1,1,0,0.1}
\definecolor{rossoCP3}{cmyk}{0,.88,.77,.40}
\definecolor{rosso}{cmyk}{0,1,1,0.4}
\definecolor{rossos}{cmyk}{0,1,1,0.55}
\definecolor{rossoc}{cmyk}{0,1,1,0.2}
\definecolor{verdes}{cmyk}{0.92,0,0.59,0.4}

\hypersetup{colorlinks, bookmarksopen, bookmarksnumbered,
citecolor=verdes, linkcolor=bluc, pdfstartview=FitH, urlcolor=rossos}

\newcommand{\mio}[1]{}

\definecolor{Gray}{gray}{0.95}

\providecommand{\abs}[1]{\lvert#1\rvert} 
\newcommand{\vect}[1]{\mathbf{#1}} 
\newcommand{\pd}[1]{\partial#1} 
\newcommand{\pr}[1]{\prime#1} 
\newcommand{\vk}[1]{ \vect{k}#1} 

\usepackage{amsmath} 
\usepackage{xcolor,graphicx} 
\usepackage{fancyvrb} 
\usepackage{caption}
\usepackage{subcaption}
\usepackage{amsfonts} 
\usepackage{color}
\usepackage{amssymb}
\usepackage{physics}
\usepackage{verbatim}
\usepackage{xfrac}
\usepackage{calc}

\usepackage{multicol}
\usepackage{color}
\definecolor{rosso}{cmyk}{0,1,1,0.4}
\definecolor{rossos}{cmyk}{0,1,1,0.55}
\definecolor{rossoc}{cmyk}{0,1,1,0.2}
\definecolor{blu}{cmyk}{1,1,0,0.3}
\definecolor{blus}{cmyk}{1,1,0,0.6}
\definecolor{bluc}{cmyk}{1,1,0,0.1}
\definecolor{verde}{cmyk}{0.92,0,0.59,0.25}
\definecolor{verdec}{cmyk}{0.92,0,0.59,0.15}
\definecolor{verdes}{cmyk}{0.92,0,0.59,0.4}

\renewcommand\&{&}

\def\circa#1{\,\raise.3ex\hbox{$#1$\kern-.75em\lower1ex\hbox{$\sim$}}\,}

\newcommand{\beq}{\begin{equation}}
\newcommand{\eeq}{\end{equation}}

\newcommand{\bea}{\begin{eqnarray}}
\newcommand{\eea}{\end{eqnarray}}
\newcommand{\be}{\begin{equation}}
\newcommand{\ee}{\end{equation}}
\newfam\rsfsfam
\def\mathscr#1{{\fam\rsfsfam\relax#1}}

\def\circa#1{\,\raise.3ex\hbox{$#1$\kern-.75em\lower1ex\hbox{$\sim$}}\,}
\makeatletter

\def\hhref#1{\href{http://arxiv.org/abs/#1}{arXiv:#1}} 

\newcommand{\doi}[1]{\href{http://dx.doi.org/#1}{[doi]}}

\def\hhref#1{\href{http://arxiv.org/abs/#1}{arXiv:#1}} 
 
\def\art{\@ifnextchar[{\eart}{\oart}}
\def\eart[#1]#2#3#4#5#6{{\rm #2}, {\em #3 \bf #4} {\rm (#6) #5} ({\em #1})}

\def\article{\@ifnextchar[{\earticle}{\oarticle}}
\def\oarticle#1#2#3#4#5#6{{\rm #1}, {\em ``#6''}, {\rm #2 #3 (#5) #4}}
\def\earticle[#1]#2#3#4#5#6#7{{\rm #2}, {\em ``#7''}, {\rm #3 #4 (#6) #5}  [\hhref{#1}]}
\def\hepart[#1]#2{{\rm #2, \em#1}}
\def\heparticle[#1]#2#3{#2, {\em ``#3''} [\hhref{#1}]}

\newcounter{alphaequation}[equation]
\def\thealphaequation{\theequation\hbox to
0.6em{\hfil\alph{alphaequation}\hfil}}
\def\eqnsystem#1{
\def\@eqnnum{{\rm (\thealphaequation)}}
\def\@@eqncr{\let\@tempa\relax \ifcase\@eqcnt \def\@tempa{& & &} \or
  \def\@tempa{& &}\or \def\@tempa{&}\fi\@tempa
  \if@eqnsw\@eqnnum\refstepcounter{alphaequation}\fi
\global\@eqnswtrue\global\@eqcnt=0\cr}
\refstepcounter{equation} \let\@currentlabel\theequation \def\@tempb{#1}
\ifx\@tempb\empty\else\label{#1}\fi
\refstepcounter{alphaequation}
\let\@currentlabel\thealphaequation
\global\@eqnswtrue\global\@eqcnt=0 \tabskip\@centering\let\\=\@eqncr
$$\halign to \displaywidth\bgroup \@eqnsel\hskip\@centering
$\displaystyle\tabskip\z@{##}$&\global\@eqcnt\@ne
\hskip2\arraycolsep\hfil${##}$\hfil& \global\@eqcnt\tw@\hskip2\arraycolsep
$\displaystyle\tabskip\z@{##}$\hfil
\tabskip\@centering&\llap{##}\tabskip\z@\cr}
\def\endeqnsystem{\@@eqncr\egroup$$\global\@ignoretrue} \makeatother

\definecolor{fiorentina}{rgb}{.5,0,.5}

\begin{document}

\vspace{1truecm}
 
\begin{center}
\boldmath

{\textbf{\Large Chiral primordial gravitational waves in extended theories of Scalar-Tensor gravity}}

\unboldmath

\unboldmath

\bigskip\bigskip

\vspace{0.1truecm}

{\bf Maria Mylova \footnote{Electronic address: M.MYLOVA.919046@swansea.ac.uk}}

{\it   Physics Department, Swansea University, Swansea, SA2 8PP, UK  }\\[1mm]

\vspace{1cm}

\thispagestyle{empty}
{\large\bf\color{blus} Abstract}
\begin{quote}

	We re-examine the problem of parity violation in single field inflation. We look for a systematic way to parametrically approach the scale at which maximal parity violation occurs, which is where we expect to find the presence of the Chern Simons instability. We do so by considering possible realizations of the effective field theory of Scalar-Tensor gravity, which could offer a rich phenomenology. The gravitational action is extended to include derivatively coupled interactions which, by means of a disformal transformation, are scaled by negative powers of a small parameter which is identified with the graviton speed. This results in suppressing the cutoff scale of the effective theory leading to parametrically large chiral tensor fluctuations. We conclude that a change in the physical description of the system is necessary in order to maintain sufficient parity violation as well ensure stability of the modes.

\end{quote}
\thispagestyle{empty}
\end{center}

\setcounter{page}{1}
\setcounter{footnote}{0}



\section{Introduction}

\indent
In Einstein's theory of General Relativity parity is conserved. On the other hand, extensions to gravity, motivated by high energy physics, require the addition of parity violating terms to the Einstein-Hilbert action \cite{Jackiw:2003pm}. These corrections create a difference in the intensities of the left and right gravitational wave polarizations resulting to a net circular polarization in the gravitational wave background. 

It has been shown that if such an asymmetry was generated during inflation it could leave an observable trace in the cosmic microwave background (CMB), i.e. by producing non-vanishing TB and EB mode correlations \cite{Lue:1998mq, Saito:2007kt, Seto:2006hf, Seto:2007tn}, whose amplitude is characterized by the scalar-tensor ratio $r$ and the degree of polarization $\Pi$. The latter is defined as the difference between the tensor power spectra of left- and right-helicity modes, at the end of inflation, normalized by the total amplitude 

\begin{equation}\begin{split}
\Pi = \frac{\mathcal{P}_h^L - \mathcal{P}_h^R}{\mathcal{P}_h^L+ \mathcal{P}_h^R},  \quad -1 < \Pi < 1,
\label{eq:pi1}
\end{split}\end{equation}
and it can take values between $-1$ and $1$ for maximally right or left-handed signal or zero when circular polarization is completely absent from the gravitational wave background.

Future experiments, such as SPIDER, CMBpol and LiteBIRD, will measure the B-Mode polarization anisotropies with expected precision $r\sim10^{-3}$ \cite{Verde:2005ff, Amblard:2006ef, Baumann:2008aq, Hazumi:2019lys}. Additionally, a direct detection of a primordial circularly polarized stochastic gravitational wave background may be possible with  space-based interferometers, such as LISA, DECIGO (Deci-Hertz) or their proposed successor BBO (Big Bang Observer) \cite{Gair:2012nm, Smith:2005mm, Kudoh:2005as, Crowder:2012ik, Smith:2016jqs,Domcke:2019zls}, as well with PTAs (Pulsar Timing Arrays) \cite{Kato:2015bye}.

It has been argued that it, ideally, requires a large circular polarization degree $\Pi$ in order to make a detection of parity violation \cite{Saito:2007kt, PhysRevD.81.123529, Nati:2017lnn} (for a more recent summary see \cite{Shandera:2019ufi}). Unfortunately, theoretical predictions for single field inflation tend to suffer quantitatively due to the Chern Simons instability \cite{Dyda:2012rj, Alexander:2004wk, Satoh:2007gn, Satoh:2008ck, Satoh:2010ep, Sorbo:2011rz}, leading to a negligible chirality enhancement. 

Alternative approaches have been to consider two-field inflationary models \cite{Satoh:2007gn}, by introducing a curvaton or several gauge fields \cite{Sorbo:2011rz}, couplings in term of a massive field \cite{Cai:2016ihp} or to consider models of inflation involving non-Abelian gauge fields \cite{Noorbala:2012fh}. The latter was an extension of the work in \cite{Alexander:2004us} where parity violating tensor perturbations, during inflation, were used to build a leptogenesis model in which they focused on the short distance modes. This model was further extended in \cite{Alexander:2018fjp} to examine baryogenesis in the dark sector.  Other approaches have been to study gravitational wave polarisation in Horava-Lifshitz gravity \cite{Takahashi:2009wc, Wang:2012fi}. See also  \cite{McDonough:2018xzh} for a top-down approach.

In this work we re-examine the problem of parity violation in the context of single field inflation. The generic effective field theory (EFT) for single field inflation was studied by Weinberg in \cite{Weinberg:2008hq} in which they produced the most general set of operators that contribute to the lower energy physics. The gravitational action was extended at next-to-leading order (NLO) to include the Weyl squared tensor and the gravitational Chern Simons term, with the latter being responsible for parity violation.  There it was shown that for the energy expansion to be finite\footnote{This is a generic effective field theory where various coefficients are assumed to be of order unity and the strength of the higher-order corrections is determined by dimensional analyses.} the heavy mass scale $\Lambda$ suppressing the higher-derivative operators cannot be much smaller than the reduced Planck mass $M_{Pl}$. Considering that $M_{Pl} \simeq 10^{18}$ GeV and the energy scale of inflation is constrained to be $H \lesssim 10^{13}$ GeV \cite{Ade:2015lrj} one would expect parity violating effects to be very small. Therefore, different assumptions are needed if one wishes to examine a regime where new physics are expected to be much closer to the scales that we can measure in the CMB.

In this work we aim to approach this regime parametrically. We ideally want to keep the heavy mass scale $\Lambda \simeq M_{Pl}$ fixed and instead introduce a parameter whose effect is to parametrically suppress the value of $\Lambda$. The most straightforward way to achieve this is by introducing a non-trivial dispersion relation. This process is well understood in, for example, the  effective field theory of inflation (EFTI) studied by Cheung et al. \cite{Cheung:2007st}. There it is the broken time diffeomorphism (St\"uckelberg trick) which introduces  extra pieces to the action. This, naturally, results to a non-trivial speed for the scalar and tensor sectors whose effect is to suppress the heavy mass scale of the non-quadratic fluctuations resulting to the strong coupling scale of the theory appearing parametrically below the scale at which the background was integrated out, leading to large scalar non-Gaussianity. Later on, they addressed this issue in \cite{Baumann:2011su, Gwyn:2012mw} by studying weakly coupled completions of the EFTI action for scalar fluctuations. $P(X)$ theories, where $X=-1/2 \pd_\mu \phi \pd^\mu \phi$, are also known to produce a similar scenario in the non-relativistic limit \cite{deRham:2017aoj}.

Here we approach this from the point of view of Horndeski theories (or beyond) which are known to be characterized by non-trivial dispersion relations. Assuming there exists a UV completion for the Horndeski theory, one could take a bottom-up approach in which the Horndeski theory could be extended, from an effective field theory point of view, by writing down the most general set of operators that agree with the symmetries and particle content of the full theory. So far only partial extensions to parity-preserving Scalar-Tensor theories have been attempted \cite{Solomon:2017nlh, deRham:2018red} while ghost-free parity violating corrections to Scalar-Tensor theories were separately examined in \cite{Crisostomi:2017ugk}.

Here we produce an extension to the Horndeski theory (or beyond) by employing a disformal transformation of the metric. It is well known that disformal transformations can generate Horndeski or beyond Horndeski theories. Special disformal transformations, where the disformal function depends only on the scalar field $\phi$,  were analysed in \cite{Bettoni:2013diz}. General disformal mappings of the Einstein-Hilbert action were considered in \cite{Zumalacarregui:2013pma}, while general disformal transformations of quadratic DHOST Lagrangians were investigated in \cite{Crisostomi:2016czh, Achour:2016rkg}. There it was shown that disformal transformations introduce extra pieces to the action which, naturally, change the dispersion relation  for gravitons.

We extend on these ideas by including disformally transformed higher-curvature operators from an EFT point of view. As one expects, the formulations can quickly grow to be too cumbersome when working with disformal transformations at the covariant level. While we have already taken steps towards that direction (as an example see Appendix \ref{appA4}), it is convenient to first examine disformal transformations of higher-curvature operators for cosmological perturbations as they are simpler. The results of this work are to be understood as indicative of what can be achieved when one considers chiral scalar-tensor extensions to Horndeski's theory and beyond. 

The aim of this work is to parametrically suppress the heavy mass scale $\Lambda$ of the higher-derivative operators. To examine this in a systematic way and in the simplest way possible, we employ an inverse disformal transformation \cite{Domenech:2015hka} on the extended action for tensors in \cite{Weinberg:2008hq}. We find that if the higher-curvature  operators are scaled by inverse powers of a small parameter $c_T$, which we identify with the graviton speed, this pushes their contributions into the  ``UV sensitive" regime, where the energy scale suppressing these corrections is well below the scale of the reduced Planck mass $M_{Pl}$, leading to \textit{parametrically} large chiral tensor fluctuations. In this sense the effective field theory is organized into an energy expansion, where $E/\Lambda_*$ is the expansion parameter and  $\Lambda_*$ is an effective mass scale which parametrically depends on some power of $c_T$. We find maximal parity violation occurs for parametrically small values of $c_T \ll 1$ in the limit that $E^2/\Lambda_*^2 \rightarrow 1$. This signals the breakdown of the perturbative expansion and the presence of the Chern Simons instability. Furthermore we expect cubic or higher-order interactions to become important resulting to a strongly coupled theory.

We conclude that a change in the physical description of the system is necessary in order to maintain sufficient parity violation as well ensure stability of the theory. Preliminary results on cubic interactions indicate that one would ideally need to work with a weakly coupled completion of our system. Such completions have already been attempted in the scalar sector of the effective field theory of inflation (EFTI) studied by Cheung et al. \cite{Baumann:2011su, Gwyn:2012mw}. 

The paper is organized as follows. In Section \ref{appS2} we introduce the quadratic action which we disformally transform in Section \ref{appS21}. We then examine the validity of the EFT in Section \ref{appS22} and proceed to produce second-order equations of motion in Section \ref{appS33}. Next, we identify the conditions for a stable Chern-Simons theory in Section \ref{appS44} and proceed to solve the linearised theory in the simplest way possible in Section \ref{appS55}. Finally we evaluate the power spectrum in Section \ref{appS66} and briefly look at the different representations of the theory in Section \ref{appS77}. We conclude our results in Section \ref{appS88}.

\section{Parametric amplification of chiral gravitational waves in single field inflation}
\label{appS2}

We consider possible realizations of the effective field theory  of Scalar-Tensor gravity which could offer a rich phenomenology. One can start with a gravity + scalar system and extend it with higher-derivative operators from an EFT point of view. Here we achieve this with an inverse disformal transformation of the quadratic action. We take a bottom-up approach, in which the action is organized as an energy expansion, where the leading-order Lagrangian is that of standard canonical Einstein gravity, while the higher-order Lagrangians are treated perturbatively. At next-to-leading order (NLO), i.e. four-derivative operators, the action reads \cite{Weinberg:2008hq}

\begin{equation}\begin{split}
S^{(0)}& = \frac{M_{Pl}^2}{2} \int \dd^4{x}  \sqrt{-  g} \Bigg\{ R+\frac{f_1}{M_{Pl}^2}  W^{\mu\nu\rho\sigma}  W_{\mu\nu\rho\sigma} 
+ \varepsilon^{\mu\nu\rho\sigma}  \frac{f_2}{M_{Pl}^2}  W_{\mu\nu\kappa\lambda}  W^{ \kappa\lambda}_{ \ \ \ \rho\sigma} \Bigg\} ,
\label{eq:action1}
\end{split}\end{equation}
where $W_{\mu\nu\rho\sigma}$ is the Weyl tensor, and $\varepsilon^{\mu\nu\rho\sigma}  = (-g)^{-\frac{1}{2}} \epsilon^{\mu\nu\rho\sigma}$ is the totally antisymmetric Levi-Civita tensor density. The reduced Planck mass is defined as $M_{Pl}^{-1} = \sqrt{8\pi G}$, with $G$ being the gravitational Newton's constant. The last term in (\ref{eq:action1}) is the gravitational Chern Simons term, which is sometimes denoted as $W \tilde  W$ and it is responsible for parity violation. 

We denote with $\Lambda\simeq M_{Pl}$ the energy scale of the heavy degrees of freedom that have been integrated out of the Lagrangian. Such high energy processes are not experimentally accessible to us, but instead they enter the low energy action,  order-by-order, though the coefficients in the derivative expansion. Therefore, higher-order corrections are expected to be subdominant as they are suppressed by the heavy mass scale $\Lambda$ which is the energy scale at which we expect to find "new physics". For a great review on EFTs see \cite{petrov2015effective, baumann2015inflation} and the excellent reviews by C.P. Burgess.

The functions $f_1(\phi)$  and $f_2(\phi)$ represent generic couplings of the dimensionless scalar field $\phi= \phi_c /\Lambda \simeq \phi_c/ M_{Pl}$ which satisfies inflationary dynamics and is homogeneous and isotropic. In the absence of a UV complete description, the form of the coupling strengths can be motivated from phenomenological considerations and/or experimental observations. 

From now on we choose to work with the conformal time $\eta$ and assume an isotropic and homogeneous FRW cosmology with line element

\begin{equation}\begin{split}
\dd{ s}^2 = a(\eta)^2 [- \dd{\eta}^2 + \dd{x}^2],
\label{eq:1aaa}
\end{split}\end{equation}
where $a = - (H \eta)^{-1}$ is the scale factor and take the scalar field to be homogeneous $\phi=\phi(t)$. Although, during inflation, the de Sitter symmetries are taken to be broken we choose, for simplicity, to work in an approximately exact de Sitter space with the Hubble parameter given by $H(t) \sim H = \text{const.}$ and where $t$ is the proper cosmic time. Finally, the metric is expanded, up to second-order in perturbations, around a de Sitter background

\begin{equation}\begin{split}
g_{\mu\nu}= \tilde{g}_{\mu\nu} + h_{\mu\nu}, \quad \frac{h_{\mu\nu}}{g_{\mu\nu}} \ll 1,
\label{eq:1aaaa}
\end{split}\end{equation}
where the perturbations respect the transverse-traceless (TT) conditions, namely $\pd_i h^i_j = h^i_i =0$. In what follows we focus only on the tensor sector. 

\subsection{Disformally transformed action}
\label{appS21}

We choose to express the quadratic action in terms of barred parameters (see Appendix \ref{appA1}). Therefore, we consider the effects of the inverse of a disformal transformation of the form

\begin{equation}\begin{split}
g_{\mu\nu} \rightarrow  \bar{g}_{\mu\nu} = c_T \qty[ g_{\mu\nu} + (1-c_T^2) n_\mu n_\nu].
\label{eq:disf5}
\end{split}\end{equation} 
Here we have used the normalization $n_\mu = \phi_{,\mu} / \sqrt{2X}$ with $n_\mu n^\mu = -1$ and $X= - \tfrac{1}{2} \phi_{,\mu} \phi^{,\mu}$. We follow the methods in \cite{Domenech:2015hka, Creminelli:2014wna, Baumann:2015xxa} where it was shown that in an FRW cosmological setting, with the scalar field $\phi$ taken to be homogeneous, a disformal transformation corresponds to a redefinition of the time-coordinate and the scale factor. In particular, for an inverse disformal transformation, one can make the following redefinitions

\begin{equation}\begin{split}   
\dd{\bar{\eta}}=  c_T \dd{\eta}, \quad \bar{a} = c_T^{\frac{1}{2}} a  ,
\label{eq:disf11}
\end{split}\end{equation}
where for simplicity we take the disformal parameter $c_T=\text{const}.$ and $\bar f_1, \bar f_2$ to be functions of the conformal time $\bar\eta$. At second-order in perturbations of the metric\footnote{ Here we used the formulations in \cite{Soda:2011am} and the Mathematica package in \cite{Pitrou:2013hga} to produce the perturbed expressions.} the contributions to (\ref{eq:action1}) transform as follows (see Appendix \ref{appA1} for more general expressions)

\begin{equation}\begin{split}
S^{(2)}&=\frac{M^2_{Pl}}{2} \int \dd^4{x} \Bigg\{ \frac{a^2}{4}\qty[(h_{ij}^\prime)^2-c_T^2 (\nabla h_{ij})^2] 
+ \epsilon^{ijk0}  \frac{8 \bar f_2}{c_T^2M_{Pl}^2} \pd_i h_{lj}^\prime \qty[\mathcal{H} h_{lk}^\prime + h_{lk}^{\prime\prime} ]
\\& +  \frac{ \bar f_1}{ M_{Pl}^2} \qty[ \frac{ (h_{ij}^{\prime\prime})^2}{ 2 c_T^3}+ \frac{h_{ij}^{\prime\prime} \nabla^2  h_{ij}}{c_T}  +\frac{c_T(\nabla^2  h_{ij})^2}{2}-\frac{2 (\nabla h_{ij}^\prime)^2}{c_T}  ] \Bigg\},
\label{eq:action}
\end{split}\end{equation} 
where we denote time derivatives with a prime ($\pd_\eta= '$). The Hubble parameter in terms of the conformal time is given by $\mathcal{H}=aH$ and we use Latin indices to denote spatial components. At leading order in (\ref{eq:action}) we find a quadratic action with gravitons having a non-trivial speed, given by

\begin{equation}\begin{split}
S^{(2)}_{LO}&=\frac{M^2_{Pl}}{2} \int \dd^4{x}  \frac{a^2}{4}\qty[(h_{ij}^\prime)^2-c_T^2 (\nabla h_{ij})^2].
\label{eq:act1}
\end{split}\end{equation} 
This is because, at the level of the perturbations, the effect of a disformal transformation can be seen as a stretching of the time-coordinate with respect to the spatial coordinate. This can also be understood by considering the effect of a disformal transformation at the covariant level.  For example, in \cite{Zumalacarregui:2013pma} (see Appendix B.1, relation (B10) in \cite{Zumalacarregui:2013pma}) it was shown that the effect of a pure disformal transformation to the Einstein-Hilbert action translates to adding extra pieces to the action which results in producing a Horndeski like theory. Such theories are known to be characterised by non-trivial dispersion relations. In this sense, the next natural step is to identify the parameter $c_T$ in (\ref{eq:action}) to be the leading order contribution to the tensor speed.  Similarly, the higher-curvature contributions are modified by acquiring extra pieces (as an example see Appendix \ref{appA4}).

It is interesting to note that the leading order action in (\ref{eq:act1}) is related to the quadratic action for tensors in \cite{Cheung:2007st}, namely

\begin{equation}\begin{split}
S^{(2)}&=\frac{M^2_{Pl}}{2} \int \dd^4{x}  \frac{a^2}{4} c_T^{-2} \qty[(h_{ij}^\prime)^2-c_T^2 (\nabla h_{ij})^2] = \frac{ \tilde M^2_{Pl}}{2} \int \dd^4{x}  \frac{a^2}{4}  \qty[(h_{ij}^\prime)^2-c_T^2 (\nabla h_{ij})^2],
\label{eq:act2}
\end{split}\end{equation} 
by a conformal transformation which can be used to set the modified Planck mass $\tilde M_{Pl}$ in (\ref{eq:act2}) to standard. In \cite{Cheung:2007st} it is a broken time diffeomorphism (St\"uckelberg trick) which introduces  extra pieces to the action. This naturally results to a non-trivial speed for gravitons. It would be interesting to see how parity violation is affected in their setup. In their work they focused on curvature perturbations which were treated as Goldstone boson modes. There it was found that a small scalar speed reduces the mass scale that suppresses the non-quadratic fluctuations leading to sizeable non-Gaussianity.

Here we find that, in a similar fashion, higher-order quadratic operators (and consequently non-quadratic operators) with more time-derivatives are enhanced, for small parameter $c_T \ll1$, compared to operators with spatial derivatives\footnote{A similar conclusion was reached in \cite{Cheung:2007st} for the scalar sector. In their case it was the spatial derivatives that were enhanced with respect to time-derivatives.}. In this way the sub-leading terms in (\ref{eq:action}) could become sizeable as they are scaled by negative powers of a small parameter, leading to \textit{parametrically large chiral tensor fluctuations}. Next we look at the phenomenological consequences of our set-up.

\subsection{Validity of the EFT}
\label{appS22}

The main aim of this work is to parametrically approach the regime at which the EFT breaks down, which is where we expect the Chern Simons instability to appear in the system.  

The effective action in (\ref{eq:action}) is organized into an energy expansion which is in terms of powers of the expansion parameter $E/\Lambda_*$, where $\Lambda_*$ is an effective mass scale proportional to some power of the parameter $c_T$. The energy expansion will continue to be valid until the sub-leading terms become as important as the leading order terms. Therefore we expect that the cosmological perturbation theory will break down in the non-relativistic limit $c_T \ll 1$, i.e our effective field theory will cease to be meaningful if the ratio $E/\Lambda_* \rightarrow 1$. Putting this into a Feynman language, at this point one would expect the propagator of the free field theory to pick up substantial contributions which can affect the leading order kinematics. This can also result into the presence of unphysical states in the system \cite{Weinberg:2008hq}.
The usual prescription is that one will need to add extra pieces to the action in order to restore the validity of the EFT\footnote{At some stage one may have to work with the UV complete description of the system.}.

To find when this happens we need to estimate the form of the effective mass scale $\Lambda_*$ by examining the action in (\ref{eq:action}) which is quadratic in the fields. The case where $c_T=1$ was considered in \cite{ArmendarizPicon:2008yv}. Here we wish to find the effective mass scale of (\ref{eq:action}) for $c_T \ll 1$.  We employ the methods in \cite{Baumann:2011su} (see Section 2 in \cite{Baumann:2011su}). We restore fake Lorentz invariance by defining the following rescaling of the spatial coordinates $x \rightarrow \tilde{x} = c_T^{-1} x$ and the canonically normalized tensor perturbation

\begin{equation}\begin{split}
\tilde{h}_{ij}^2 = \frac{ c_T^3 M_{Pl}^2  h_{ij}^2}{4}.
\label{eq:st1}
\end{split}\end{equation} 
From this we find that the action in (\ref{eq:action}) becomes

\begin{equation}\begin{split}
S^{(2)}&=\frac{1}{2} \int \dd{\eta} \dd^3{\tilde x} \Bigg\{ \qty[(\tilde h_{ij}^\prime)^2- ( \tilde \nabla \tilde h_{ij})^2] 
+ \epsilon^{ijk0}  \frac{32 \bar f_2}{c_T^3M_{Pl}^2 } \tilde{\partial}_i \tilde h_{lj}^\prime \qty[\mathcal{H} \tilde h_{lk}^\prime +\tilde h_{lk}^{\prime\prime} ]
\\& +  \frac{ \bar f_1}{ c_T^3 M_{Pl}^2 } \qty[ 2(\tilde h_{ij}^{\prime\prime})^2+ 4 \tilde h_{ij}^{\prime\prime} \tilde \nabla^2 \tilde h_{ij}  +2(\tilde \nabla^2  \tilde h_{ij})^2- 8 (\tilde \nabla \tilde h_{ij}^\prime)^2  ] \Bigg\}.
\label{eq:st2}
\end{split}\end{equation} 
The effective mass scale of the theory $\Lambda_*^2 \simeq c_T^3 M_{Pl}^2$ can be read off directly from (\ref{eq:st2}), where for simplicity we  treat $\bar f_1$ and $\bar f_2$ as order-one parameters. To ensure the validity of low energy observables we need our perturbative expansion to hold at the relevant scales that we can measure in the CMB, i.e. at horizon crossing where the fluctuations freeze. Therefore, we demand that the size of the fluctuations at the de Sitter scale $k_{ph}\sim H$, where $k_{ph} = k/a$ is the physical momentum, coming from the higher-order corrections, is much less than $\mathcal{O}(1)$. Therefore, we have

\begin{equation}\begin{split}
\frac{ k_{ph}^2}{ \Lambda_*^2} \simeq \frac{ H^2}{c_T^3 M_{Pl}^2} \ll 1,
\label{eq:st3}
\end{split}\end{equation} 
which translates to a lower bound on the graviton speed. 


Additionally, the need for a finite perturbative expansion may require us to include next-to-next-to-leading order (NNLO) operators, i.e. six-derivative terms, as they can pick up enhancements that could stand them relevant to the calculation. This is because at (NNLO) the energy expansion ratio will be of order $H^4/\Lambda_*^4$ (see Appendix \ref{appA3}). Higher-derivative parity preserving extensions to scalar-tensor gravity were discussed in \cite{Solomon:2017nlh} while higher-derivative parity-violating operators for Scalar-Tensor chiral theories were discussed in \cite{Crisostomi:2017ugk}, where alongside the gravitational Chern Simons term they included first- and second-derivatives of the scalar field. These were subsequently studied in \cite{Qiao:2019hkz}. We only tentatively look at these in Appendix \ref{appA3}. 

Additionally, in the limit $c_T \ll1$  non-quadratic terms can become important i.e. terms of the form $\sim h h h $. Parity violation in tensor non-Gaussianity was investigated in \cite{Maldacena:2011nz, Soda:2011am}. They showed that there is no parity violation in de Sitter, but found non-vanishing contributions to the bispectrum when slow-roll inflation is taken into account. In particular the three-point correlators $\langle TTB \rangle$, $\langle TEB \rangle$, $\langle EEB \rangle$ could become non-vanishing, as opposed to the parity conserving case, which we could potentially observe in the CMB data. Furthermore, mixed correlators were more recently studied in \cite{Bartolo:2017szm}.

\begin{figure}[htbp]
	\begin{center}
		\includegraphics[width=0.6\textwidth]{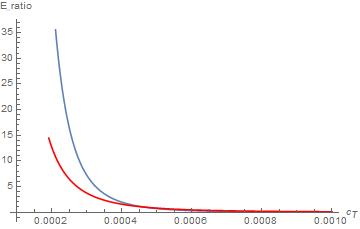}
		\caption{Once the parameter $c_T$ attains the value in (\ref{eq:st3d}) cubic interactions (in blue) become more important than the quadratic ones (in red), signalling the break down of the effective description. For this plot we take $M_{Pl}= 10^{18}$ GeV and $H=10^{13}$ GeV. The horizontal axis displays the values of $c_T$ in the interval $\{10^{-4}, 10^{-3}\}$ which is the range in which we expect to find the value for $c_T$ in (\ref{eq:st3d}) for this particular numerical example, while the vertical axis displays the energy expansion ratio at the non-relativistic limit $ E_{ratio} \sim \mathcal{O}(1)$.}
		\label{fig:2a}
	\end{center}
	\setlength{\abovecaptionskip}{0pt plus 1pt minus 1pt}
\end{figure}

We do not consider in detail the effect of higher-order interaction terms in this work although it would be interesting to examine how the bispectrum is affected in a more realistic scenario of our setup. In terms of our arguments, we require that cubic interactions remain sub-leading. As an example, take a cubic interaction of the form\footnote{Such interaction terms can be found by expanding the Weyl squared tensor to third-order in perturbations of the metric.}


\begin{equation}\begin{split}
S^{(3)}&=\frac{M^2_{Pl}}{2} \int \dd{\eta} \dd^3{ x} \frac{ f_1}{M_{Pl}^2} \qty[\frac{1}{c_T^3} \mathcal{H} h_{ij}^\prime h_{il}^\prime h_{lj}^\prime - 2 c_T  h_{ij} \pd_i \pd_j h^{lm} \nabla^2 h_{lm} + \cdots ]
\label{eq:st3b}
\end{split}\end{equation}
Once we restore fake Lorentz invariance, we find

\begin{equation}\begin{split}
S^{(3)}&=\frac{1}{2} \int \dd{\eta} \dd^3{ x} c_T^{-\frac{9}{2}} \frac{ f_1}{  M_{Pl}^3 } \qty[8 \mathcal{H} h_{ij}^\prime h_{il}^\prime h_{lj}^\prime - 16   h_{ij} \pd_i \pd_j h^{lm} \nabla^2 h_{lm} + \cdots ].
\label{eq:st3c}
\end{split}\end{equation}
From this we can see that the energy expansion ratio is of the order $E^3 / \Lambda_*^3$. We can approximate the following constraint coming from demanding that the cubic interactions do not dominate at around the energy scale of inflation $k_{ph} \sim H$, which is

\begin{equation}\begin{split}
\frac{ k_{ph}^3}{ \Lambda_*^3} \simeq c_T^{-\frac{9}{2}} \frac{H^3}{ M_{Pl}^3} \ll 1.
\label{eq:st3a}
\end{split}\end{equation} 
We see that in the limit $c_T \ll 1$ new physics could appear at energies not far above the energy scale of inflation. Consequently, the higher-order interactions may acquire large couplings which could leave measurable evidence, from the new physics, in the CMB data. Furthermore, the Chern Simons instability will appear in our system for some $\mathcal{O}(1)$ value of (\ref{eq:st3}) and (\ref{eq:st3a}) for which we find the limiting value

\begin{equation}\begin{split}
c_T \simeq \qty(\frac{H}{M_{Pl}})^\frac{2}{3}.
\label{eq:st3d}
\end{split}\end{equation} 
We will reach the same conclusions when we examine the effective potential, later on in the text. The behaviour of the coefficients for the quadratic and cubic interactions around the limit (\ref{eq:st3d}) is shown in Figure (\ref{fig:2a}). From this we conclude that the Chern Simons instability is a consequence of the strongly coupled theory. This can be seen in the graph, as $c_T$ approaches the limit in (\ref{eq:st3d}) the cubic interactions (in blue) shoot up with respect to the quadratic interactions (in red) which, by definition, leads to the theory being strongly coupled. 

\subsection{The equations of motion }
\label{appS33}

Before we begin with our analysis, we need to briefly discuss the Ostrogradsky instability whose no-go theorem \cite{Ostrogradsky:1850fid, Woodard:2015zca} is central in the study of higher-order corrections to gravity. The effective Lagrangian in (\ref{eq:action}) includes contributions that lead to higher than second-order equations of motion, which cannot always be removed by partial integration. While we cannot excite the Ostrogradsky ghost, as long as we remain in the low-energy regime of the EFT, the system can still exhibit unphysical effects if the equations of motion are higher than order-two\footnote{It requires additional initial conditions in order to eliminate unwanted solutions. At best such systems can only be solved numerically.}. 

For (NLO) contributions to the action it is possible to obtain second-order equations of motion via a field redefinition, which amounts to a substitution in terms of the equations of motion obtained from the leading order action in (\ref{eq:act1})

\begin{equation}\begin{split}
h_{ij}^{\prime\prime}+2\mathcal{H}h_{ij}^\prime-c_T^2 \nabla^2 h_{ij}=0.
\label{eq:dr1}
\end{split}\end{equation}
This way one can shift the offending terms at higher-orders in the expansion. As one expects, this can complicate things when working at (NNLO). It is not, in general, trivial to find field redefinitions that ensure second-order equations of motion when the action contains a combination of (NLO) and (NNLO) operators.  Although, such field redefinitions may be possible to find, they tend to be quite involved. As an example, we would like to point the reader to the analysis that was performed in \cite{Gong:2014rna} for the scalar sector. It would be interesting to see if, in the future, we could obtain something similar for tensors.  As we do not wish to enter into lengthy discussions regarding these issues we refer the reader to \cite{Georgi:1991ch, GrosseKnetter:1993td, Arzt:1993gz} and references therein (see also \cite{Manohar:2018aog} for a recent pedagogical treatment). 

For these reasons, here we focus at (NLO) corrections to gravity and only briefly discuss (NNLO) operators in Appendix \ref{appA3}. At second-order in perturbations of the metric, we expect the disformally transformed action to contain, schematically, the following type of contributions

\begin{equation}\begin{split}
S^{(2)}&=\frac{M^2_{Pl}}{8} \int \dd^3{x} \dd{\eta}  a^2 \Bigg\{ \qty(1+ \alpha) (h_{ij}^\pr)^2
-\qty(c_T^2+\beta) (\nabla  h_{ij})^2
-\gamma (\nabla h_{ij}^\pr)^2 + \delta  (\nabla^2 h_{ij})^2 
\\&- \epsilon^{ijk}\qty[\varepsilon (h^q_{ \ i})^\pr(\pd_j h_{kq})^\pr- \zeta  (\pd^r h^q_{ \ i})\pd_j\pd_r h_{kq}] \Bigg\},
\label{eq:1b01}
\end{split}\end{equation} 
where the coefficients $\alpha$ - $\zeta$ are functions of the conformal time. This form of the action guarantees second-order equations of motion. Using (\ref{eq:dr1}) we find that the action in (\ref{eq:action}) takes the following form

\begin{equation}\begin{split}
S^{(2)}&=\frac{M^2_{Pl}}{8} \int \dd^3{x} \dd{\eta}  a^2 \Bigg\{ \qty(1+\frac{ \bar f_1 \mathcal{H}^2}{c_T^3 a^2  M_{Pl}^2})  (h_{ij}^\pr)^2 -\qty[c_T^2+\frac{\qty(\bar f_1^\prime \mathcal{H} + \bar f_1\mathcal{H}^\prime)}{c_T a^2  M_{Pl}^2}  ] \qty(\nabla h_{ij})^2
\\& -\frac{\bar f_1}{c_T a^2  M_{Pl}^2}  \qty(\nabla h_{ij}^\prime)^2 + \frac{\bar f_1 c_T}{a^2 M_{Pl}^2}  \qty(\nabla^2 h_{ij})^2 
- \frac{ \bar f_2^\prime}{a^2 M_{Pl}^2}  \epsilon^{ijk}\qty[ \frac{1}{c_T^2} h_{qi}^\pr \pd_j h_{kq}^\pr-(\pd^r h^q_{ \ i})\pd_j\pd_r h_{kq}] \Bigg\},
\label{eq:modEFT1}
\end{split}\end{equation} 
where we have absorbed numerical factors into the definitions of $\bar f_1$ and $\bar f_2$. It is easy to see that (\ref{eq:modEFT1}) has the same form with (\ref{eq:1b01}) which guarantees second-order equations of motion. Next, we produce the Euler-Lagrange equations which are expressed in terms of the canonically normalized field\footnote{Here we use the conventions in \cite{Baumann:2009ds}.}  $\mu^s_{k} = (M_{Pl}/2)   h^s_{k}  z^s_{k}$, which read as (see Appendix \ref{appA1})

\begin{equation}\begin{split}
(\mu^s_k)^{\prime\prime} + \qty[ k^2 \frac{\qty(c_T^2+\frac{ ( \bar f_1 \mathcal{H}^\prime+\bar f_1^\prime \mathcal{H})}{c_T  a^2 M_{Pl}^2}-\frac{ c_T k^2 \bar f_1}{ a^2 M_{Pl}^2} -\frac{ \lambda^s k \bar f_2^\prime}{ a^2 M_{Pl}^2} )}{\qty(1+ \frac{  \bar f_1 \mathcal{H}^2}{c_T^3 a^2 M_{Pl}^2}-\frac{ k^2 \bar f_1}{c_T  a^2 M_{Pl}^2} -\frac{ \lambda^s k \bar f_2^\prime}{c_T^2  a^2 M_{Pl}^2})}- \frac{(z^s_k)^{\prime\prime}}{z^s_k} ] \mu^s_k = 0,
\label{eq:311}
\end{split}\end{equation}
where the effective potential is defined in terms of

\begin{equation}\begin{split}
z^s_k  = a \sqrt{1+ \frac{  \bar f_1 \mathcal{H}^2}{c_T^3 a^2 M_{Pl}^2}-\frac{ k^2 \bar f_1}{c_T  a^2 M_{Pl}^2} -\frac{ \lambda^s k \bar f_2^\prime}{c_T^2 a^2 M_{Pl}^2}}.
\label{eq:z1}
\end{split}\end{equation}
The parameter $s=L,R$ is used to denote left and right graviton modes and we have defined $\lambda^s= \pm 1$. Finally, we have omitted summation over left and right modes. It is now evident that the effective potential depends on the polarization modes. This produces an asymmetry in the amplitude of the solutions for left and right modes which leads to a circularly polarized gravitational wave background \cite{Saito:2007kt}. Next we look at the Chern Simons instability.

\section{On the stability of parity breaking theories}
\label{appS44}

Let us for a moment simplify our arguments by setting $c_T=1$ in (\ref{eq:z1}). It is convenient to express the effective potential in terms of the Chern Simons and Weyl squared tensor dynamical mass-scales \cite{Dyda:2012rj} by defining  

\begin{equation}\begin{split}
M_{cs} = \frac{a M_{Pl}^2 }{ f_2^\prime},  \qq{and} M_w^2 =  \frac{M_{Pl}^2}{ f_1}, 
\label{eq:11aa}
\end{split}\end{equation}
respectively, where $k_{ph} = k /a $ is the physical wavenumber and as we will soon find out, $M_w^2 < 0$. Therefore, (\ref{eq:z1}) becomes

\begin{equation}\begin{split}
z^s_k =a  \sqrt{1+\frac{\mathcal{H}^2}{ a^2 M_{w}^2} - \frac{k_{ph}^2}{ M_{w}^2  }- \frac{k_{ph} \lambda^s}{ M_{cs} }  }.
\label{eq:11a}
\end{split}\end{equation}
If we ignore the Weyl squared tensor contributions, for a moment, it is then straight forward to see that the linear theory will break down for $(z^s_k)^2 =0$, i.e. at $k_{ph}=M_{cs}$. The amplitude of one of the helicity mode develops an instability which appears as a logarithmic divergence \cite{Alexander:2004wk}. In what follows we derive a constraint that can ensure the stability of all modes within the regime of the validity of the EFT. For the stability of the solutions it requires

\begin{equation}\begin{split}
(z^s_k)^2  > 0 \Rightarrow \frac{4}{M_{w}^2} + \frac{4 \mathcal{H}^2}{ a^2 M_{w}^4} < - \qty(\frac{\lambda^s}{M_{cs}})^2.
\label{eq:12a}
\end{split}\end{equation}
The quantity on the RHS of the inequality is negative. This implies that the only way to satisfy this relationship is to demand that the Weyl squared dynamical mass scale is tachyonic with

\begin{equation}\begin{split}
f_1<0.
\label{eq:10aa}
\end{split}\end{equation}
This unfortunately implies the loss of perturbative unitarity due to positivity bounds requiring $f_1>0$ \cite{Cheung:2016wjt}. We can now try to simplify relation (\ref{eq:12a}) using that $\mathcal{H}^2 a^{-2} = H^2$ in de Sitter. In the next step, we make it explicit that the inequality is satisfied by substituting for $M_w^2 = - \abs{M_w^2}$. Solving for $M_{cs}$, gives the simple relation

\begin{equation}\begin{split}
M_{cs}^2 > \frac{ \abs{M_w}^4}{4(\abs{M_{w}}^2-4H^2)}.
\label{eq:15a}
\end{split}\end{equation}
As this involves dynamical quantities, in the more general cases, it translates to a constraint on the time-evolution of the theory. But there is a special case, which we would like to emphasize, where the dynamical contributions completely drop out. For specific choices\footnote{The functions $f_1$ and $f_2$ have to depend on time in such a way so that all time-variables exactly cancel in (\ref{eq:15a}). We give such an example later in the text.}  of the functions  $f_1$ and $f_2$ one finds a simple relationship between the physical scales involved\footnote{For simplicity, here we have neglected a small constant contribution to the leading order term.}, namely $ \Lambda^2 > H^2$. Notice that this relationship is always satisfied in EFTs as the scale where we expect to find new physics is always larger than the energy scale of inflation (i.e. in our case we have $\Lambda \simeq M_{Pl} > H^2$). A similar conclusion was reached in \cite{Satoh:2008ck} in the case of slow-roll inflation where dynamical parameters depend only weakly on time.


\section{A toy model}
\label{appS55}

We proceed to solve the equations of motion in (\ref{eq:311}).  We choose to work in pure de Sitter ($a=-(H\eta)^{-1}$) and look at a particularly simple example, where\footnote{Here we have reinstated the numerical factors that we had absorbed earlier into the functions $ f_1$ and $ f_2$.} $\abs{f_1} = 8 \abs{f_{10}}$ and

\begin{equation}\begin{split}
f_2= 8 f_{20}\int \eta^{-1} \dd{\eta},
\label{eq:14}
\end{split}\end{equation}
with the coupling constants having magnitudes of $\abs{f_{10}}, f_{20} \sim\mathcal{O}(1) \sim 1$. This way we have that $\bar f_1= f_1$ and $\bar f_2= f_2$. We introduce the \textit{relative} parameters  $\omega_{1}, \omega_2, g_1$ and $g_2$. These are defined as follows:

\begin{equation}\begin{split}
\omega_1= \frac{ f_1}{c_T}, \quad \omega_2 = f_1 c_T, \quad g_1 = \frac{ f_2}{c_T^2}, \quad g_2= f_2.
\label{eq:2b}
\end{split}\end{equation}
This way the relationships between the functions $\omega_{1}, \omega_2, g_1$, and $g_2$, simplify to

\begin{equation}\begin{split}
\omega_1= \frac{\omega_2}{c_T^2} \qq{and}  g_1 = \frac{g_2}{c_T^2}.
\label{eq:3b}
\end{split}\end{equation}
We can make contact with (\ref{eq:1b01}) by making the  following identifications $ \gamma= \omega_1 (a M_{Pl})^{-2},\delta = \omega_2 (a M_{Pl})^{-2}, \varepsilon = g_1^\prime (a M_{Pl})^{-2}$ and $ \zeta = g_2^\prime (a M_{Pl})^{-2}$. Here we have set $\alpha=\beta=0$ as, with our definitions,  these terms only add a negligible constant to the leading order term. The equations of motion simplify to

\begin{equation}\begin{split}
(\mu^s_k)^{\prime\prime} + \qty[ k^2 \frac{\qty(c_T^2-\frac{ k^2 \omega_2}{ a^2 M_{Pl}^2} -\frac{ \lambda^s k g_2^\prime}{ a^2 M_{Pl}^2} )}{\qty(1-\frac{ k^2 \omega_1}{ a^2 M_{Pl}^2} -\frac{ \lambda^s k g_1^\prime}{  a^2 M_{Pl}^2})}- \frac{(z^s_k)^{\prime\prime}}{z^s_k} ] \mu^s_k = 0.
\label{eq:eom8}
\end{split}\end{equation}
The speed of gravitons contains corrections coming from the higher-order operators which, as we shall soon see, they drop out. The effective potential is given in terms of

\begin{equation}\begin{split}
z^s_{k}  =  a  \sqrt{1  -\frac{ k^2 \omega_1}{  a^2 M_{Pl}^2} -\frac{ \lambda^s k g_1^\prime}{  a^2 M_{Pl}^2}}.  
\label{eq:eom9}
\end{split}\end{equation}
To ensure stability for all modes, within the regime of validity of the EFT, we derive the simple constraint

\begin{equation}\begin{split}
(z^s_{k})^2 > 0 \Rightarrow M_{cs}^2 > \frac{\abs{M_{w}}^2}{4c_T^3 }, \quad  f_1 <0,
\label{eq:9b}
\end{split}\end{equation}
which can be re-expressed as a lower bound to the speed of gravitons

\begin{equation}\begin{split}
c_T^3 > \frac{\abs{M_w}^2}{4 M_{cs}^2} \simeq \frac{H^2 }{ M_{Pl}^2}.
\label{eq:10b}
\end{split}\end{equation}
Comparing this to (\ref{eq:st3}) it is now evident that the Chern Simons instability will kick in for parametrically small values of $c_T$, in the limit $E^2/\Lambda_*^2 \rightarrow 1$. 
With all this in place we can now treat the higher-order corrections as being energetically negligible for as long as we remain within the regime of validity of the EFT. From now on we make explicit the minus sign in $f_1=-\abs{f_1} <0$,  as demanded by (\ref{eq:9b}) and, for simplicity, drop the absolute notation. The effective potential is expressed as

\begin{equation}\begin{split}
\frac{(z^s_k)^{\prime\prime}}{z^s_k}&=\frac{2}{\eta^2}-\frac{1}{\eta}\qty(\frac{2{\omega_{10}} k^2\eta-g_{10} \lambda^s k}{1+{\omega_{10}} k^2\eta^2-g_{10}\lambda^s k \eta})
\\&-\frac{1}{4}\qty(\frac{2 {\omega_{10}} k^2\eta-g_{10} \lambda^s k}{1+ {\omega_{10}}{}k^2\eta^2-g_{10}\lambda^s k \eta})^2
+\qty(\frac{\omega_{10}k^2}{1+{\omega_{10}} k^2\eta^2-g_{10}\lambda^s k \eta}),
\label{eq:15}
\end{split}\end{equation}
where we have simplified our arguments by introducing the following redefinitions

\begin{equation}\begin{split}
\omega_{10} = \frac{8 f_{10} H^2}{c_T M_{Pl}^2}, \quad \omega_{20} = \frac{8 f_{10} c_T H^2}{ M_{Pl}^2}, \quad g_{10} = \frac{8 f_{20} H^2}{c_T^2 M_{Pl}^2}, \quad g_{20} = \frac{8 f_{20} H^2}{ M_{Pl}^2}.
\label{eq:16}
\end{split}\end{equation}
Using the definitions given above it is easy to see that the corrections to the leading order contribution to the speed of gravitons drop out, giving

\begin{equation}\begin{split}
\tilde{c}_T^2=\frac{c_T^2 + {\omega_{20}}{}\,k^2\eta^2 -g_{20}\lambda^s k \eta}{1+ {\omega_{10}} k^2\eta^2 -g_{10}\lambda^s k \eta} = c_T^2.
\label{eq:17}
\end{split}\end{equation}
Note that if we set $\omega_{10}= 0$ and $c_T=1$ in (\ref{eq:15}) we obtain the form of the effective potential in \cite{Alexander:2004wk}. Similarly, if we set $g_{10} = 0$ we  correctly recover the equation for infllation which corresponds to a harmonic oscillator with a time-dependent frequency. Taking the small scale limit ($\abs{k\eta} \rightarrow \infty$) of the equation of motion gives 

\begin{equation}\begin{split}
(\mu^s_k)^{\prime\prime} + k^2 c_T^2 \mu^s_k=0.
\label{eq:18}
\end{split}\end{equation}
This is satisfied by the Bunch-Davies vacuum solution where gravitons propagate with a speed $c_T$, given by

\begin{equation}\begin{split}
\lim_{\abs{k \eta} \rightarrow \infty} \mu^s_k (\eta)= - \frac{ e^{\displaystyle -i c_T k \eta}}{\sqrt{2c_T k}}.
\label{eq:19}
\end{split}\end{equation}
Taking the large scale limit ($\abs{k\eta} \rightarrow 0$), gives 

\begin{equation}\begin{split}
(\mu^s_k)^{\prime\prime} + \qty(k^2 c_T^2 - \frac{2}{\eta^2} - \frac{g_{10} k \lambda^s}{\eta} ) \mu^s_k=0,
\label{eq:20}
\end{split}\end{equation}
where, for simplicity, we have neglected subdominant contributions to the graviton speed. Equation (\ref{eq:20}) can be solved exactly. We bring it into the Whittaker form 

\begin{equation}\begin{split}
(\mu^s_{k})^{\prime\prime}  (\chi) + \qty[-\frac{1}{4} + \frac{\nu}{ \chi} + \frac{\frac{1}{4} - \mu^2}{\chi^2}]\mu^s_{k}  (\chi) =0,
\label{eq:22b}
\end{split}\end{equation}
by making a substitution of the form 

\begin{equation}\begin{split}
\chi&=2i c_T k \eta.
\label{eq:21}
\end{split}\end{equation}
Hence, equation to solve is

\begin{equation}\begin{split}
(\mu^s_{k})^{\prime\prime}  (\chi) + \qty[-\frac{1}{4} + \frac{ig_{10} \lambda^s}{2c_T \chi} + \frac{(-2)}{\chi^2}]\mu^s_{k}  (\chi) =0.
\label{eq:22b1}
\end{split}\end{equation}
Solutions are in terms of Whittaker functions\footnote{ The prescription for finding inflationary solutions in terms of the Whittaker equations can be found in \cite{Alexander:2004wk, Satoh:2010ep, Martin:2000ei}.} (the details can be found in Appendix \ref{appA2}). The tensor power spectrum, per polarization, is found to be

\begin{equation}\begin{split}
k^3 P^s_h (k) = \frac{1}{\pi^2} \frac{H^2}{M_{Pl}^2 c_T^3} \abs{\Gamma\qty(2-\frac{ig_{10}\lambda^s}{2c_T})}^{-2} \,e^{-\frac{\pi \lambda^sg_{10}}{2c_T}},
\label{eq:23}
\end{split}\end{equation}
and it is scale-invariant, as expected. We conclude that, for $\lambda=-1$, the left modes are enhanced.

\subsection{Chirality enhancement}
\label{appS66}

In this work we are interested to parametrically approach the regime at which maximal parity violation occurs. Therefore we consider the non-relativistic limit $c_T \ll 1$. To find the maximum possible enhancement for the left modes we have to take into account the condition in (\ref{eq:st3}) which is equivalent to the lower bound for $c_T$ in (\ref{eq:10b}). We find that as we parametrically approach the limit $E^2/\Lambda_*^2 \rightarrow 1$ the exponential argument in (\ref{eq:23}), namely 

\begin{equation}\begin{split}
e^{\displaystyle\frac{\pi \ g_{10}}{2c_T}} \simeq e^{\displaystyle\frac{4\pi H^2}{ c_T^3 M_{Pl}^2}}\simeq e^{\displaystyle \frac{H^2}{\Lambda^2_*}},
\label{eq:32b1}
\end{split}\end{equation}
becomes of order one\footnote{ Here we used relation (\ref{eq:st3}) to substitute for $c_T^{-3} \sim M_{Pl}^2 / H^2$.} leading to maximal parity violation with circular polarization ratio $\Pi \rightarrow 1$, indicating a fully left-handed signal. This is the point at which we expect the quadratic theory to break down and the Chern Simons instability to appear in our system. If we were to include cubic interactions we would also expect the theory, at this point, to be strongly coupled resulting to large parity violation in tensor non-Gaussianity.  

At this point we would like to emphasize that by using this framework it becomes explicit that the Chern Simons instability is a consequence of the strongly coupled theory.  From this we conclude that if maximal parity violation is to occur new physics are ought to be included. This translates to adding extra degrees of freedom so that we can admit a weakly coupled description of our system which could enable us to consistently parametrize it when we extrapolate it to higher energies. Such approaches have been attempted in the scalar sector in \cite{Baumann:2011su, Gwyn:2012mw} where they studied weakly coupled completions of the EFTI action in \cite{Cheung:2007st}. It would be interesting to do something similar for the tensor sector. 



Our conclusions are in agreement with the literature, as so far we have not been able to produce an observable amount of circular polarization of gravitational waves in single field inflation.  Our approach is to be understood as being complimentary to previously examined cases in the literature. 

Another important constraint may come from requiring a small scalar-tensor ratio as per the results from the Planck collaboration \cite{Ade:2015lrj}. Looking at the tensor power spectrum in (\ref{eq:23}) we see there is an inverse factor of $c_T^3$ scaling the overall power spectrum, therefore, when we enhance chirality at the same time we also enhance the overall amplitude of the gravitational waves, so care must be taken. Taking again the limit $E^2/\Lambda_*^2 \rightarrow 1$ and summing over the polarizations we find the power spectrum can take the maximal value\footnote{ For simplicity, here we have ignored various order one parameters.}

\begin{equation}\begin{split}
k^3 P^s_h (k) \sim  \mathcal{O}(1),
\end{split}\end{equation}
leading to a large tensor amplitude. Either way,  if the scalar-tensor ratio is too small then it would be very difficult to detect circular polarization due to cosmic variance. Therefore, a large tensor amplitude is preferable. The presence of $c_T^{-3}$ in the power spectrum could be used to parametrically enhance the amplitude of gravitational waves which also enhances detectability.

\section{Relation between frames} 
\label{appS77}

It has been shown that  disformal transformations cannot remove four- or higher-derivative corrections to the quadratic action for tensors \cite{Bordin:2017hal}. Additionally, sufficiently complicated theories do not guarantee to have an Einstein frame. For example, consider the scalar-tensor action in (\ref{eq:1b01}) for arbitrary parameters $\alpha - \zeta$. Upon disformal transformation, in the new frame, one may expect to find a quadratic action in terms of canonical Einstein gravity (i.e. as in \cite{Creminelli:2014wna}) but with a non-standard higher-curvature extension. As here we work with theories that go beyond Einstein's gravity, we do not technically consider physics in the Einstein frame. Therefore, in what follows we simply dub different frames as A and B. 

In our case, we started in a frame, say A, where the action in (\ref{eq:action1}) is described by the canonical Einstein quadratic term plus the extension to gravity given by Weinberg in \cite{Weinberg:2008hq}. This theory is understood to be valid up to some fixed heavy mass scale $\Lambda \simeq M_{Pl}$.  We then moved our formulations to another frame, say B, by disformally mapping the  operators at the level of the perturbations. We found the action in (\ref{eq:action}), in which gravitons propagate with a non-trivial speed $c_T$. We demonstrated that if $c_T$ becomes parametrically small for a fixed scale $\Lambda\simeq M_{Pl}$ it could spoil the validity of the EFT.  We showed that this is the point where we expect maximal parity violation to occur resulting to the presence of the Chern Simons instability. Additionally, cubic operators will become important resulting into tensor modes appearing strongly coupled well below the energy scale $\Lambda \simeq M_{Pl}$.

It is to be understood that we work, essentially, with a different representation of extensions to Einstein's gravity. In this sense, the A and B frame theories are said to be mathematically equivalent \cite{Sotiriou:2007zu}. In the limit where $c_T=1$ the theories are also said to be physically equivalent. On the other hand, when we consider arguments on the physical equivalence between frames, at the non-relativistic limit $c_T \ll 1$, we stumble across a noticeable difference.  We cannot bypass the effective field theory machinery which instructs us that  we should have to include additional degrees of freedom to the B frame theory, so that we can produce a theory that is valid and consequently weakly coupled all the way from the energy scale of inflation $H$ to the fixed cutoff scale $\Lambda \simeq M_{Pl}$. This is very likely to affect the effective description of the theory (as in \cite{Baumann:2011su,  Gwyn:2012mw}) and therefore, the observables as they are sensitive to the higher-curvature corrections. We find that arguments on physical equivalence between frames are more difficult to reconcile once we go beyond leading order effects. 

Finally, let us remark that one could build an effective description having the same form with (\ref{eq:action}) by starting with operators that do not admit an inverse disformal transformation. These were examined in \cite{Chagoya:2016inc}.

\section{Outlook}
\label{appS88}

In this work we re-examined the problem of parity violation in single field inflation. In particular, we looked for a systematic way to parametrically approach the scale at which maximal parity violation occurs. We achieved this by introducing a small parameter, by means of a disformal transformation, whose effect was to suppress the heavy cutoff scale $\Lambda \simeq M_{Pl}$ of the effective theory leading to parametrically large chiral tensor fluctuations. 

We found that sub-leading quadratic operators can become important signalling the presence of the Chern Simons instability. This inevitably implies the existence of non-trivial cubic interactions which could stand the theory strongly coupled for parametrically small values of the parameter $c_T$ and, consequently, lead to large parity violation in tensor non-Gaussianity. We showed that by using this framework it becomes explicit that the Chern Simons instability is a consequence of the strongly coupled theory. 

We concluded that, at this point, a change in the physical description of the theory is necessary so that one can consistently parametrize our system as the energy increases up to the heavy cutoff scale $\Lambda \simeq M_{Pl}$. The addition of new degrees of freedom could help, in a consistent manner, to maintain sufficient parity violation while at the same time ensure stability of the modes. We hope to examine this in a future work. 


Our approach is to be understood as being complimentary to previously examined cases in the literature. In particular, extra field content could be incorporated into our EFT framework in a systematic way.


At this point we would like to emphasize that our work is only indicative of the many open questions one needs to carefully tackle when considering higher-order extensions to Scalar-Tensor gravity, especially in the presence of parity violation. We do not wish to answer these questions in one go. As a first attempt we kept the formulations as simple as possible and only tentatively looked at these problems. Indeed, more complicated effective descriptions can come from considering disformal transformations of operators at the covariant level. Below we address the possible directions one can take so that progress can be made in the future.  

The next step would be to find a phenomenologically viable inflationary model and do a complete and concrete analysis in quasi-de Sitter. Additionally, one could take into account the complete set of non-redundant six-derivative operators, as higher-curvature terms can become important. Such considerations are subject to finding a field redefinition that can ensure second-order equations of motion. It would also be interesting to see how the bispectrum is affected in this setup by considering cubic operators. In particular we are interested in the regime in which these contributions become important, leading to large tensor non-Gaussianity.

Finally, at the end of inflation the scalar field has reached its minimum ($\phi = \text{const}.$) and the gravitational Chern Simons term becomes a total derivative. Therefore, we do not expect to find parity violation in the subsequent evolution of the classical gravitational dynamics \cite{Satoh:2008ck}. On the other hand, the six-derivative parity violating terms may not necessarily become surface terms which may affect post-inflationary predictions. This deserves more investigation.

\section*{Acknowledgements}

We are grateful to Gianmassimo Tasinato, Carlos N\'u\~nez, Gonzalo A. Palma and Ivonne Zavala Carrasco for very useful discussions and comments on the manuscript. We would like to thank Miguel Zumalac\'arregui for providing the code which was used as basis for the computations in Appendix \ref{appA4}. The calculations in Section \ref{appS55} were done in collaboration with G. Tasinato. MM is supported by an STFC studentship under the DTP ST/N504464/1.

\appendix

\section{Disformal transformation of linearised (NLO) operators}
\label{appA1}

Our starting point is the, most familiar to us, extension to Einstein's gravity as discussed in \cite{Weinberg:2008hq}. At the background level the action is given by 

\begin{equation}\begin{split}
S&=\frac{M^2_{Pl}}{2}  \int \dd^4{x} \sqrt{-g} \qty[ R + \frac{f_1}{\Lambda^2} W^{\mu\nu\rho\sigma} W_{\mu\nu\rho\sigma} + \frac{f_2}{\Lambda^2} \epsilon^{\mu\nu\rho\sigma} g^{\kappa\beta} g^{\lambda\zeta} W_{\mu\nu\beta\zeta} W_{\rho\sigma\kappa\lambda} ].
\label{eq:0004}
\end{split}\end{equation} 
This is expanded around an FRW background to second-order in perturbations of the metric. We choose to express the perturbed action in term of barred parameters. The action reads

\begin{equation}\begin{split}
S^{(2)}&= \int \dd^3{x} \dd{\bar{\eta}}  \Bigg\{ \frac{M^2_{Pl}}{8} \bar{a}^2 \qty[(\pd_{\bar{\eta}} h_{ij})^2 - (\pd_k h_{ij})^2]
\\&+ \frac{M^2_{Pl}}{4\Lambda^2} \bar{f_1} \qty[\qty(\pd_{\bar{\eta}}(\pd_{\bar{\eta}} h_{ij}) + \nabla^2 h_{ij} )^2  - 4 (\pd_k \pd_{\bar{\eta}} h_{ij})^2]
\\&- \frac{M^2_{Pl}}{2\Lambda^2} 8 \bar{f}_2 \epsilon^{ijk} (\pd_{\bar{\eta}}\pd_i h_{lj})  \qty[  \bar{\mathcal{H}} (\pd_{\bar{\eta}}h_{lk})  + \pd_{\bar{\eta}}(\pd_{\bar{\eta}} h_{lk})] \Bigg\},
\label{eq:0001}
\end{split}\end{equation} 
where $\bar f_1$ and $\bar f_2$ are dimensionless functions of time. We introduce modifications to gravity by performing an inverse disformal transformation following the prescription in \cite{Domenech:2015hka, Creminelli:2014wna, Baumann:2015xxa}.

When dealing with only (NLO) operators one can employ a suitable field redefinition to ensure second-order equations of motion. There are two ways to do this here. One is to first transform the action and then use the equations of motion from the leading order contributions to the quadratic action in the new frame to reduce the order of the time derivatives. Equivalently one can use the leading order equations of motion in the barred frame to reduce the order of the time derivatives  and then transform the action to the new frame. The result will be the same. In this Appendix we choose the latter, therefore, using the equations of motion for the quadratic action in the barred frame, namely

\begin{equation}\begin{split} 
\pd_{\bar{\eta}}(\pd_{\bar{\eta}} h_{ij}) + 2\bar{\mathcal{H}} \pd_{\bar{\eta}} h_{ij} - \nabla^2 h_{ij}=0,
\label{eq:0002}
\end{split}\end{equation}
the action reduces to

\begin{equation}\begin{split}
S^{(2)}&= \int \dd^3{x} \dd{\bar{\eta}}  \Bigg\{\frac{M^2_{Pl}}{8}   \bar{a}^2 \qty[(\pd_{\bar{\eta}} h_{ij})^2 - (\pd_k h_{ij})^2]
\\&+ \frac{M^2_{Pl}}{\Lambda^2}  \qty[\bar f_1 \bar{\mathcal{H}}^2  (\pd_{\bar{\eta}} h_{ij})^2
-\qty[ (\pd_{\bar{\eta}} \bar f_1) \bar{\mathcal{H}}+\bar f_1 (\pd_{\bar{\eta}}\bar{\mathcal{H}})] (\nabla h_{ij})^2
-\bar f_1 (\pd_{\bar{\eta}}\nabla h_{ij})^2  + \bar f_1  (\nabla^2h_{ij}  )^2]
\\&- \frac{M^2_{Pl}  }{\Lambda^2} (\pd_{\bar{\eta}} \bar f_2)  \epsilon^{ijk}\qty[  (\pd_{\bar{\eta}}h_{q i}) (\pd_{\bar{\eta}}\pd_j h_{kq}) + (\nabla^2 h_{ q i})\pd_j h_{kq}] \Bigg\}.
\label{eq:0003}
\end{split}\end{equation} 
We choose the following redefinition of the time-coordinate and the scale factor 

\begin{equation}\begin{split}
\dd{\bar{\eta}} = c_T \dd{\eta}, \quad \bar{a} = c_T^{-\frac{1}{2}} \mathcal{F}^{\frac{1}{2}} a,
\label{eq:disf1}
\end{split}\end{equation} 
where $c_T= \mathcal{F} \mathcal{G}^{-1}$ and with the parameters  $\mathcal{F}, \mathcal{G}$ being functions of time. We find the Hubble parameter transforms as  

\begin{equation}\begin{split}
\bar{\mathcal{H}} = c_T^{-1} \qty(\mathcal{H} - \frac{1}{2} \frac{c_T^\prime}{c_T} + \frac{1}{2} \frac{\mathcal{F}^\prime}{\mathcal{F}}).
\label{eq:disfhubble}
\end{split}\end{equation} 
In the new frame the action becomes

\begin{equation}\begin{split}
S^{(2)}&=\frac{M^2_{Pl}}{8} \int \dd^3{x} \dd{\eta}  a^2 \Bigg\{ \qty(\mathcal{G}+\frac{8 \bar f_1 \mathcal{A}}{c_T^3 a^2  \Lambda^2})  (h_{ij}^\pr)^2 -\qty[\mathcal{F}+\frac{8\qty(\bar f_1^\prime \mathcal{B}_1 + \bar f_1\mathcal{B}_2)}{c_T a^2  \Lambda^2}  ] \qty(\nabla h_{ij})^2
\\& -\frac{8\bar f_1}{c_T a^2  \Lambda^2}  \qty(\nabla h_{ij}^\prime)^2 + \frac{8\bar f_1 c_T}{a^2 \Lambda^2}  \qty(\nabla^2 h_{ij})^2 
\\&- \frac{8 \bar f_2^\prime}{a^2\Lambda^2}  \epsilon^{ijk}\qty[ \frac{1}{c_T^2} h_{qi}^\pr \pd_j h_{kq}^\pr-(\pd^r h^q_{ \ i})\pd_j\pd_r h_{kq}] \Bigg\},
\label{eq:modEFT}
\end{split}\end{equation} 
where we denote derivatives with respect to the conformal as $'=\pd_{\eta}$ and have defined the following parameters

\begin{equation}\begin{split}
&\mathcal{A} = \mathcal{H}^2 + \mathcal{H} \qty(\frac{\mathcal{F}^\prime}{\mathcal{F}}-\frac{c_T^\prime}{c_T}) + \qty(\frac{1}{2} \frac{\mathcal{F}^\prime}{\mathcal{F}}- \frac{1}{2}\frac{c_T^\prime}{c_T})^2,
\\& \mathcal{B}_1 =\mathcal{H} + \frac{1}{2} \frac{\mathcal{F}^\prime}{\mathcal{F}}- \frac{1}{2}\frac{c_T^\prime}{c_T}, \quad 
\mathcal{B}_2 = \mathcal{H}^\prime -\mathcal{H}\frac{c_T^\prime}{c_T} - \frac{1}{2} \frac{c_T^\prime}{c_T} \frac{\mathcal{F}^\prime}{\mathcal{F}} + \qty(\frac{c_T^\prime}{c_T})^2 - \frac{1}{2}\qty( \frac{\mathcal{F}^\prime}{\mathcal{F}})^2 + \frac{1}{2} \frac{\mathcal{F}^{\prime\prime}}{\mathcal{F}}- \frac{1}{2}\frac{c_T^{\prime\prime}}{c_T}. 
\label{eq:modEFTparam}
\end{split}\end{equation} 
We do not know yet how the functions $\bar f_1$ and $\bar f_2$ transform. At the limit of constant parameter $c_T$ the parameters $\mathcal{A}, \mathcal{B}_1$ and $ \mathcal{B}_2$ reduce to 

\begin{equation}\begin{split}
\mathcal{A}=  \mathcal{H}^2 +  \mathcal{H}\frac{\mathcal{F}^\prime}{\mathcal{F}}+ \frac{1}{4} \qty(\frac{\mathcal{F}^\prime}{\mathcal{F}})^2 , \quad \mathcal{B}_1 =\mathcal{H} + \frac{1}{2} \frac{\mathcal{F}^\prime}{\mathcal{F}}, \quad \mathcal{B}_2 = \mathcal{H}^\prime  - \frac{1}{2}\qty( \frac{\mathcal{F}^\prime}{\mathcal{F}})^2 + \frac{1}{2} \frac{\mathcal{F}^{\prime\prime}}{\mathcal{F}}.
\label{eq:modEFTparam1}
\end{split}\end{equation} 
In the case where $\mathcal{F},\mathcal{G}$ are constants we simply have

\begin{equation}\begin{split}
\mathcal{A}=  \mathcal{H}^2 , \quad \mathcal{B}_1 =\mathcal{H}, \quad \mathcal{B}_2 = \mathcal{H}^\prime, 
\label{eq:modEFTparam2}
\end{split}\end{equation} 
i.e. the Hubble parameter simply transforms as a derivative $\bar{ \mathcal{H}} = c_T^{-1}  \mathcal{H}$. We proceed to produce the equations of motion. 
The Fourier transform for tensor perturbations is given by

\begin{equation}\begin{split}
h_{ij}(\eta,\vect{x})=\frac{1}{(2\pi)^{\frac{3}{2}}} \int \dd{\vect{k}} \sum_{s=R,L} p^s_{ij}({\vect{k}}) h^s_{k}(\eta) e^{\displaystyle i \vect{k}\cdot \vect{x}},
\label{eq:eom4cs}
\end{split}\end{equation} 
where the tensor polarizations are defined in a circular basis as 

\begin{equation}\begin{split}
p^R_{ij} \equiv \frac{1}{\sqrt{2}} (p^+_{ij} + i p^\times_{ij}) \qq{and} p^L_{ij} \equiv \frac{1}{\sqrt{2}} (p^+_{ij} - i p^\times_{ij}) = (p^R_{ij})^*,
\label{eq:pol3}
\end{split}\end{equation}
where $p^+_{ij}, p^\times_{ij} $ are the two linear polarization tensors and $p^R_{ij}, p^L_{ij} $ are polarizations that rotate in the right ($R$) and left ($L$) handed directions, respectively. These satisfy the following transversality and traceless conditions

\begin{equation}\begin{split}
p^s_{ij} k_j =0 \qq{and} (p^{i}_i)^s =0, \quad s=L,R,
\label{eq:pol8}
\end{split}\end{equation}
where $\vect{k}=k_i$ are momenta in the spatial directions. In momentum space the Euler-Lagrange equations read

\begin{equation}\begin{split}
& \  \ \ \  ( h^s_k)^{\prime\prime} \qty( 1 + \frac{8 \bar f_1 \mathcal{A}}{c_T^3 \mathcal{G} a^2  \Lambda^2} - \frac{8 k^2 \bar f_1 }{c_T \mathcal{G} a^2  \Lambda^2} -\frac{8 k \lambda^s \bar f_2^\prime}{c_T^2 \mathcal{G} a^2  \Lambda^2})
\\&+ (h^s_k)^\prime \bigg( 2 \mathcal{H}  + \frac{8\bar f_1 \mathcal{A}^\prime}{c_T^3 \mathcal{G} a^2  \Lambda^2} - \frac{24\bar f_1 \mathcal{A} c_T^\prime}{c_T^4 \mathcal{G} a^2  \Lambda^2} + \frac{\mathcal{G}^\prime}{\mathcal{G}} + \frac{8\bar f_1^\prime \mathcal{A}}{c_T^3 \mathcal{G} a^2  \Lambda^2}
\\& - \frac{8  k^2 \bar f_1^\prime}{c_T \mathcal{G} a^2  \Lambda^2} + \frac{8  k^2 \bar f_1 c_T^\prime}{c_T^2 \mathcal{G} a^2  \Lambda^2} -  \frac{8 k \lambda^s \bar f_2^{\prime\prime}}{c_T^2 \mathcal{G} a^2  \Lambda^2} +  \frac{16 k \lambda^s \bar f_2^\prime c_T^\prime}{c_T^3 \mathcal{G} a^2  \Lambda^2} \bigg)
\\& + k^2 h^s_k \qty(c_T^2 +  \frac{8\bar f_1^\prime \mathcal{B}_1}{c_T \mathcal{G} a^2  \Lambda^2} +  \frac{8\bar f_1 \mathcal{B}_2}{c_T \mathcal{G} a^2  \Lambda^2}-  \frac{8 k^2\bar f_1 c_T}{ \mathcal{G} a^2  \Lambda^2} -  \frac{8 k \lambda^s \bar f_2^\prime}{ \mathcal{G} a^2  \Lambda^2}) =0,
\label{eq:modEOM2}
\end{split}\end{equation} 
where we have used the following identity to simplify the result

\begin{equation}\begin{split}
i \frac{k_p}{k} \epsilon^{pjk} p_{ik} = - \lambda^s ( p^j_i)^s, \quad \lambda^s=\pm 1,
\label{eq:eom6}
\end{split}\end{equation} 
and have omitted summation over left and right modes. Next, for simplicity we consider the case where $c_T = const.$ and $\mathcal{F}, \mathcal{G}$ are functions of time, as in (\ref{eq:modEFTparam1}).  We can obtain the evolution equations for tensor fluctuations in momentum space in terms of the canonically normalized amplitudes 

\begin{equation}\begin{split}
\mu^s_{k}(\eta) = \frac{M_{Pl}}{2 }  h^s_{k}(\eta)  z^s_{k}(\eta).
\label{eq:eom5cs}
\end{split}\end{equation} 
We find

\begin{equation}\begin{split}
(\mu^s_k)^{\prime\prime} + \qty[ k^2 \frac{\qty(c_T^2+\frac{8 ( \bar f_1 \mathcal{B}_2+\bar f_1^\prime \mathcal{B}_1)}{c_T \mathcal{G} a^2 \Lambda^2}-\frac{8 c_T k^2 \bar f_1}{\mathcal{G} a^2 \Lambda^2} -\frac{8 \lambda^s k \bar f_2^\prime}{\mathcal{G} a^2 \Lambda^2} )}{\qty(1+ \frac{ 8 \bar f_1 \mathcal{A}}{c_T^3 \mathcal{G} a^2 \Lambda^2}-\frac{8 k^2 \bar f_1}{c_T \mathcal{G} a^2 \Lambda^2} -\frac{8 \lambda^s k \bar f_2^\prime}{c_T^2 \mathcal{G} a^2 \Lambda^2})}- \frac{(z^s_k)^{\prime\prime}}{z^s_k} ] \mu^s_k = 0,
\label{eq:31}
\end{split}\end{equation}
where the effective potential is defined in terms of

\begin{equation}\begin{split}
z^s_k  = z_1 (\tau) \sqrt{1+ \frac{ 8 \bar f_1 \mathcal{A}}{c_T^3 \mathcal{G} a^2 \Lambda^2}-\frac{8 k^2 \bar f_1}{c_T \mathcal{G} a^2 \Lambda^2} -\frac{8 \lambda^s k \bar f_2^\prime}{c_T^2 \mathcal{G} a^2 \Lambda^2}}, \quad z_1(\tau) = z_0  a \sqrt{2\mathcal{G}c_T^3} \Lambda,
\label{eq:z}
\end{split}\end{equation}
and it is given by

\begin{equation}\begin{split}
\frac{(z^s_k)^{\prime\prime}}{z^s_k}  &= \frac{z_1^{\prime\prime}}{z_1} + \frac{z_1^\prime}{z_1} \qty[\frac{\qty(\frac{8 \bar f_1 \mathcal{A}}{c_T^3\mathcal{G} a^2 \Lambda^2})^\prime-k^2 \qty(\frac{8\bar f_1}{c_T\mathcal{G} a^2 \Lambda^2})^\prime - k \lambda^s \qty(\frac{8 \bar f_2^\prime}{c_T^2\mathcal{G} a^2 \Lambda^2})^\prime}{1+ \frac{ 8 \bar f_1 \mathcal{A}}{c_T^3 \mathcal{G} a^2 \Lambda^2}-\frac{8 k^2 \bar f_1}{c_T \mathcal{G} a^2 \Lambda^2} -\frac{8 \lambda^s k \bar f_2^\prime}{c_T^2 \mathcal{G} a^2 \Lambda^2}} ]
\\&+ \frac{1}{2} \qty[ \frac{\qty(\frac{8 \bar f_1 \mathcal{A}}{c_T^3\mathcal{G} a^2 \Lambda^2})^{\prime\prime}-k^2 \qty(\frac{8\bar f_1}{c_T\mathcal{G} a^2 \Lambda^2})^{\prime\prime} - k \lambda^s \qty(\frac{8 \bar f_2^\prime}{c_T^2\mathcal{G} a^2 \Lambda^2})^{\prime\prime}}{1+ \frac{ 8 \bar f_1 \mathcal{A}}{c_T^3 \mathcal{G} a^2 \Lambda^2}-\frac{8 k^2 \bar f_1}{c_T \mathcal{G} a^2 \Lambda^2} -\frac{8 \lambda^s k \bar f_2^\prime}{c_T^2 \mathcal{G} a^2 \Lambda^2}}]
\\& - \frac{1}{4} \qty[ \frac{\qty(\frac{8 \bar f_1 \mathcal{A}}{c_T^3\mathcal{G} a^2 \Lambda^2})^\prime-k^2 \qty(\frac{8\bar f_1}{c_T\mathcal{G} a^2 \Lambda^2})^\prime - k \lambda^s \qty(\frac{8 \bar f_2^\prime}{c_T^2\mathcal{G} a^2 \Lambda^2})^\prime }{1+ \frac{ 8 \bar f_1 \mathcal{A}}{c_T^3 \mathcal{G} a^2 \Lambda^2}-\frac{8 k^2 \bar f_1}{c_T \mathcal{G} a^2 \Lambda^2} -\frac{8 \lambda^s k \bar f_2^\prime}{c_T^2 \mathcal{G} a^2 \Lambda^2}}]^2,
\label{eq:31a}
\end{split}\end{equation}
with

\begin{equation}\begin{split}
\frac{z_1^{\prime\prime}}{z_1} =  \frac{a^{\prime\prime}}{a}+ \frac{1}{2}  \frac{\mathcal{G}^{\prime\prime}}{\mathcal{G}} + \frac{a^\prime}{a} \frac{\mathcal{G}^\prime}{\mathcal{G}} - \frac{1}{4} \qty( \frac{\mathcal{G}^\prime}{\mathcal{G}})^2, \quad \frac{z_1^\prime}{z_1} = \frac{a^\prime}{a} + \frac{1}{2}  \frac{\mathcal{G}^\prime}{\mathcal{G}}. 
\label{eq:31b}
\end{split}\end{equation}
To ensure correct signs for the kinetic term and a healthy speed for gravitons we define the following constraints

\begin{equation}\begin{split}
& (z^s_k)^2  > 0 \Rightarrow  1+ \frac{ 8 \bar f_1 \mathcal{A}}{c_T^3 \mathcal{G} a^2 \Lambda^2}-\frac{8 k^2 \bar f_1}{c_T \mathcal{G} a^2 \Lambda^2} -\frac{8 \lambda^s k \bar f_2^\prime}{c_T^2 \mathcal{G} a^2 \Lambda^2} >0, 
\\&\qq{and}   c_T^2+\frac{8 ( \bar f_1 \mathcal{B}_2+\bar f_1^\prime \mathcal{B}_1)}{c_T \mathcal{G} a^2 \Lambda^2}-\frac{8 c_T k^2 \bar f_1}{\mathcal{G} a^2 \Lambda^2} -\frac{8 \lambda^s k \bar f_2^\prime}{\mathcal{G} a^2 \Lambda^2} >0.
\label{eq:cst1}
\end{split}\end{equation}

\section{The equations of motion} 
\label{appA2}
The equation to solve is

\begin{equation}\begin{split}
(\mu^s_{k})^{\prime\prime}  (\chi) + \qty[-\frac{1}{4} + \frac{ig_{10} \lambda^s}{2c_T \chi} + \frac{(-2)}{\chi^2}]\mu^s_{k}  (\chi) =0,
\end{split}\end{equation}
where

\begin{equation}\begin{split}
\chi&=2i c_T k \eta.
\end{split}\end{equation}
Solutions are in terms of Whittaker functions. Here we keep the Whittaker $W$ function which has a growing solution for left modes and a negative decaying solution for right modes. The Whittaker $W$ solution is expressed as in \cite{jeffrey2007table} (page 1024, relation 9.220, 2)

\begin{equation}\begin{split}
W_{\nu,\mu}(\chi)= \chi^{\mu+\frac{1}{2}} e^{-\frac{\chi}{2}} U\qty(\mu-\nu+\frac{1}{2}, 2\mu+1;\chi),
\label{eq:23b}
\end{split}\end{equation}
where $U$ is the confluent hypergeometric function. In our case we have that

\begin{equation}\begin{split}
\nu= \frac{ig_{10} \lambda^s}{2c_T }, 
\label{eq:24b}
\end{split}\end{equation}
and $\mu$ is found to be $\frac{1}{4} - \mu^2 = -2 \Rightarrow \mu^2 = (2 + \frac{1}{4}) = \frac{9}{4} \Rightarrow \mu = \pm \frac{3}{2} $. Any root works, therefore, we choose

\begin{equation}\begin{split}
\mu=  \frac{3}{2}.
\label{eq:25b}
\end{split}\end{equation}
The full solution is given by 

\begin{equation}\begin{split}
\mu^s_{\vect{k}}  (\chi) &= B_1(k) W_{\nu,\mu}(\chi)
=  B_1(k)   e^{-i c_T k \eta} (2i c_T k \eta)^2 U\qty(2- \frac{ig_{10} \lambda^s}{2c_T } , 4; 2i c_T k \eta),
\label{eq:26b}
\end{split}\end{equation}
which is normalized against the Bunch-Davies vacuum, as follows. The asymptotic representation for the Whittaker function for large values of $\abs{\chi}$ is given by (see \cite{jeffrey2007table} page 1026, relation 9.227)

\begin{equation}\begin{split}
\lim_{\abs{\chi} \rightarrow \infty} W_{\nu,\mu}(\chi) \sim e^{-\frac{\chi}{2}} \chi^\nu.
\label{eq:27b}
\end{split}\end{equation}
Therefore, taking the small scale limit of the Whittaker function gives 

\begin{equation}\begin{split}
\lim_{\abs{k \eta} \rightarrow \infty} \mu^s_{\vect{k}}  (\chi) &= B_1(k) \lim_{\abs{k \eta} \rightarrow \infty} W_{\nu,\mu}(\chi) = B_1(k) e^{\displaystyle -i c_T k \eta } \qty(i c_T k \eta )^{\displaystyle  \frac{ig_{10} \lambda^s}{2c_T } }
\\&= B_1(k) e^{\displaystyle -i c_T k \eta }e^{\displaystyle \qty[  \frac{ig_{10} \lambda^s}{2c_T } \ln\qty(2i c_T k \eta )]}
\\&=B_1(k) e^{\displaystyle -i c_T k \eta }e^{\displaystyle \Bigg\{ \frac{ig_{10} \lambda^s}{2c_T }  \qty[\ln\qty(2 c_T\abs{k\eta})-i\frac{\pi}{2}]\Bigg\}}
\\&= B_1(k) e^{\displaystyle -i c_T k \eta}e^{\displaystyle \qty( \frac{\pi g_{10} \lambda^s}{4c_T } )},
\label{eq:28b}
\end{split}\end{equation}
up to a phase. To correctly normalize the solutions we need to compare the above to the Bunch-Davies vacuum solution (\ref{eq:19}). We have

\begin{equation}\begin{split}
\lim_{\abs{k \eta} \rightarrow \infty} \mu^s_\vk(\eta)= - \frac{ e^{\displaystyle -i c_T k \eta}}{\sqrt{2c_T k}} =  B_1(k) e^{\displaystyle -i c_T k \eta}e^{\displaystyle \qty( \frac{\pi g_{10} \lambda^s}{4c_T } )}.
\label{eq:29b}
\end{split}\end{equation}
Finally, the normalization is 

\begin{equation}\begin{split}
B1(k) = - \frac{1}{\sqrt{2c_T k}} e^{\displaystyle \qty(- \frac{\pi g_{10} \lambda^s}{4c_T } )}.
\label{eq:30b}
\end{split}\end{equation}
Substituting for $B_1(k)$ into (\ref{eq:26b}) we find that the correctly normalized Whittaker solutions are given by

\begin{equation}\begin{split}
\mu^s_k (\chi) = (-2 c_T k \eta)^{\frac{3}{2}} \sqrt{-\eta} e^{-\frac{\pi \lambda^s g_{10}}{4 c_T}} U\qty(2- \frac{i \lambda^s g_{10}}{2 c_T} , 4, 2ic_T k \eta).
\label{eq:31b}
\end{split}\end{equation}
To find the form of the solution at large scales we use (see \cite{abramowitz1965handbook}, page 508, relation 13.5.6) $\lim_{z\rightarrow 0} U(a,b,z)= \frac{\Gamma(b-1)}{\Gamma(a)} z^{1-b}, \mathcal{R} b\geq 2, b\neq 2$, giving

\begin{equation}\begin{split}
\lim_{\abs{k\eta}\rightarrow 0} \mu^s_k (\chi) =  \sqrt{\frac{-\eta}{2(-c_T k \eta)^3}} \frac{1}{\Gamma\qty(2- \frac{i g_{10} \lambda^s}{2 c_T})} e^{-\frac{\pi \lambda^s g_{10}}{4 c_T}}.
\label{eq:32b}
\end{split}\end{equation}
We need to find the power spectrum in terms of $h^s_{k}=2 \mu^s_{k} ( z^s_{k}  M_{Pl})^{-1}$. We first need to solve for

\begin{equation}\begin{split}
\frac{(z^s_{k})^{\prime\prime}}{z^s_{k}} = \frac{2}{\eta^2} + \frac{g_{10}k \lambda^s}{\eta},
\label{eq:33b}
\end{split}\end{equation}
which has the general form (see \cite{olver2010nist}, relation 2.8.24)

\begin{equation}\begin{split}
\frac{W^{\prime\prime}}{W} =  \frac{u^2}{4\xi}+\frac{(\nu^2-1)}{4\xi^2} + \frac{\psi(\xi)}{\xi},
\label{eq:33b1}
\end{split}\end{equation}
with $\nu=\pm 3, \psi(\xi) =0$. We pick the growing solution (for the left modes) to be

\begin{equation}\begin{split}
z^s_{k}(k\eta)= 2 z_0 \sqrt{g_{10} \lambda^s k\eta} K_3 (2 \sqrt{g_{10} \lambda^s k \eta}).
\label{eq:35b}
\end{split}\end{equation}
We use $\lim_{z\rightarrow0} K_n (z) = \frac{1}{2} \Gamma(n) (\frac{1}{2} z )^{-n}$ to take the large scale limit giving 

\begin{equation}\begin{split}
\lim_{\abs{k\eta}\rightarrow 0} z^s_{k}= z_0 2 \sqrt{g_{10} \lambda^s k\eta} (\sqrt{g_{10} \lambda^sk\eta} )^{-3} =-( H\eta)^{-1}.
\label{eq:36b}
\end{split}\end{equation}
The tensor power spectrum, per polarization, is found to be 

\begin{equation}\begin{split}
P^s_h (k) = \frac{1}{\pi^2} \frac{H^2}{M_{Pl}^2 c_T^3} \abs{\Gamma\qty(2-\frac{ig_{10}\lambda^s}{2c_T})}^{-2} \,e^{-\frac{\pi \lambda^sg_{10}}{2c_T}},
\end{split}\end{equation}
where for $\lambda=-1$ the left modes are enhanced.

\section{Disformal transformation of covariant (NLO) operators} 
\label{appA4}

We consider a pure disformal transformation of the higher-curvature operators in \cite{Weinberg:2008hq}. Here we follow the procedure in the Appendix of \cite{Zumalacarregui:2013pma}. Using the following field redefinition 

\begin{equation}\begin{split}
\pi=\int \dd{\phi} \sqrt{B(\phi)}, 
\label{eq:d2a}
\end{split}\end{equation}
the barred metric can be expressed in terms of the following pure disformal relations

\begin{equation}\begin{split}
\bar{g}_{\mu\nu} = g_{\mu\nu}+\pi_{,\mu}\pi_{,\nu}, \quad \bar{g}^{\mu\nu} = g^{\mu\nu} - \gamma_0^2 \pi^{,\mu}\pi^{,\nu},
\label{eq:d2b}
\end{split}\end{equation} 
with

\begin{equation}\begin{split}
\gamma_0^2=\frac{1}{1-2X_\pi}, \quad \nabla_\mu \gamma_0 = - \gamma_0^3 \pi^{,\alpha} \pi_{;\alpha\mu}, \quad X=-\frac{1}{2} \pi_{\mu} \pi^{,\mu},
\label{eq:d2c}
\end{split}\end{equation} 
where we use the notation $\pi_{,\mu} = \pd_\mu \pi$ and $\pi^{;\mu}_{ \ \ \mu} = \nabla_\mu \nabla^\mu \pi$. The action to transform was given in (\ref{eq:action1}). The transformation of the Einstein-Hilbert term was already considered in \cite{Zumalacarregui:2013pma}. Next we consider the transformation of the Weyl squared tensor which we express as

\begin{equation}\begin{split}
\bar{W}_{\mu\nu\rho\sigma} \bar{W}^{\mu\nu\rho\sigma}&=  \bar{R}_{ \mu \nu\rho\sigma} \bar{R}^{ \mu \nu\rho\sigma}-2 \bar{R}_{\mu\nu}\bar{R}^{\mu\nu}+\frac{1}{3}\bar{R}^2.
\label{eq:d15}
\end{split}\end{equation}
We find the following contributions, where we denote the Weyl squared tensor as $W^2$ for short

\begin{equation}\begin{split}
\bar{W}^2&= W_{\alpha\beta\delta\epsilon} W^{\alpha\beta\delta\epsilon}
\\&+4 \gamma_0^2 \bigg\{ R_{\alpha\delta\beta\epsilon} \Pi^{\beta\alpha} \Pi^{\epsilon\delta} -\langle R_\alpha^{ \ \delta\epsilon\zeta} R_{\beta\delta\epsilon\zeta} \rangle + \langle R^{\delta\epsilon} R_{\alpha\delta\beta\epsilon} \rangle + R^{\alpha\beta} \Pi_{\delta\beta} \Pi^\delta_{ \ \alpha}
\\& + \langle R_\alpha^{ \ \delta} R_{\beta\delta} \rangle - \frac{1}{3} R \langle R_{\alpha\beta} \rangle - [\Pi] R^{\alpha\beta} \Pi + \frac{1}{6} R \qty([\Pi]^2 -[\Pi^2])\bigg\}
\\&+4 \gamma_0^4 \bigg\{\frac{1}{2} \langle\!\langle R_{\alpha \ \beta}^{ \ \zeta \ \eta}  R_{\delta\zeta\epsilon\eta}\rangle\!\rangle - \langle R_{\alpha\epsilon\beta\zeta} \Pi^{\epsilon\delta}\Pi^{\zeta}_{ \ \delta} \rangle +  \langle R_{\alpha\epsilon\beta\zeta} \Pi^{\zeta\epsilon} \rangle 
\\&- 4 \langle R_{\beta\epsilon\delta\zeta}  \Pi^\delta_{ \ \alpha} \Pi^{\zeta\epsilon} \rangle - 2 \langle R_\alpha^{ \ \delta}  \Pi_{\epsilon\delta} \Pi^{\epsilon\beta} \rangle + \frac{1}{3}\langle R_{\alpha\beta}\rangle  \qty([\Pi^2] -[\Pi]^2  ) 
\\&+2 \langle R_\alpha^{ \ \delta} \Pi_{\delta\beta} \rangle [\Pi] -\langle R^{\delta\epsilon} \Pi_{\delta\alpha} \Pi_{\epsilon\beta} \rangle + \langle \Pi \rangle R^{\epsilon\delta} \Pi - \frac{1}{6} \langle\!\langle R_{\alpha\beta} R_{\delta\epsilon} \rangle\!\rangle
\\&+ \frac{1}{3} R \qty(\langle\Pi^2\rangle - \langle \Pi \rangle [\Pi]) + \frac{1}{12}[\Pi]^4 - \frac{2}{3} [\Pi]^2 [\Pi^2] + \frac{7}{12} [\Pi^2]^2 + [\Pi] [\Pi^3] - [\Pi^4] \bigg\}
\\&+ 4 \gamma_0^6 \bigg\{ \langle\Pi\rangle \langle R_{\delta\zeta\epsilon\eta} \Pi^{\eta\zeta} \rangle - \langle\!\langle R_{\delta\zeta\epsilon\eta} \Pi^\zeta_{ \ \alpha} \Pi^\eta_{ \ \beta} \rangle\!\rangle + \frac{1}{3}\langle R{\alpha\beta}\rangle \qty(\langle \Pi \rangle^2 -\langle \Pi \rangle [\Pi] )   
\\& + 4 \langle \Pi^4 \rangle -3 \langle \Pi^3 \rangle [\Pi] + \frac{4}{3} \langle \Pi^2 \rangle [\Pi]^2 - \frac{1}{3} \langle \Pi \rangle [\Pi]^3 -\frac{7}{3} \langle \Pi^2 \rangle [\Pi^2] 
\\&+ \frac{4}{3} \langle \Pi \rangle [\Pi] [\Pi^2] - \langle \Pi \rangle [\Pi^3] \bigg\}
\\&+ \gamma_0^8 \bigg\{\frac{4}{3}\qty(\langle\Pi^2\rangle^2+\langle\Pi\rangle \langle\Pi^2\rangle [\Pi] ) - \frac{2}{3} \langle\Pi\rangle^2 [\Pi]^2 -4\langle\Pi\rangle \langle\Pi^3\rangle +2 \langle\Pi\rangle^2 [\Pi^2] \bigg\}.
\label{eq:d16}
\end{split}\end{equation}
Here we have condensed our expressions by using the Galileon notation in \cite{Zumalacarregui:2013pma}. The Chern Simons gravitational term $W\tilde{W}$ can be expressed as

\begin{equation}\begin{split}
\bar{W\tilde{W}}= - \bar{\varepsilon}^{\mu\nu\rho\sigma} \bar R^\lambda_{ \ \kappa\rho\sigma} \bar R^\kappa_{ \ \lambda\mu\nu}-4 \bar{\varepsilon}^{\lambda\mu\nu\rho} \bar R_{\kappa\lambda} \bar R^\kappa_{\mu\nu\rho} + \frac{2}{3} \bar{\varepsilon}^{\kappa\lambda\mu\nu} \bar R \bar R_{\kappa\lambda\mu\nu},
\label{eq:d17}
\end{split}\end{equation}
where $\bar{\varepsilon}^{\mu\nu\rho\sigma}= \epsilon^{\mu\nu\rho\sigma} / \sqrt{\abs{\bar{g}}}$ is the Levi-Civita tensor density. The above transforms as

\begin{equation}\begin{split}
\bar{W\tilde{W}} &=  \bar{\varepsilon}^{\mu\nu\rho\sigma} W_{\mu\nu\kappa\lambda} W^{\kappa\lambda}_{ \ \ \rho\sigma}
\\&+4  \gamma_0^2 \bigg\{\frac{1}{6}\bar{\varepsilon}^{\mu\nu\rho\sigma} R_{\mu\nu\rho\sigma} \qty([\Pi]^2-[\Pi^2]) - \frac{1}{2} \langle \bar{\varepsilon}^{\nu\rho\sigma\alpha} R_{ \kappa \mu\nu\rho} R^{\lambda\mu}_{ \ \ \sigma\alpha} \rangle
\\& + \langle \bar{\varepsilon}^{\mu\nu\rho\sigma} R_{\kappa\mu} R_{\lambda\nu\rho\sigma}  \rangle + \langle \bar{\varepsilon}^{\nu\rho\sigma\alpha} R_{\kappa\mu\lambda\nu} R^\mu_{ \ \rho\sigma\alpha} \rangle - \frac{1}{3} \langle \bar{\varepsilon}^{\mu\nu\rho\sigma} R_{\kappa\lambda} R_{\mu\nu\rho\sigma} \rangle
\\& +  \bar{\varepsilon}^{\mu\nu\rho\sigma} R_{\lambda\nu\rho\sigma} \Pi^{\lambda\kappa} \Pi_{\mu\kappa} - [\Pi] \bar{\varepsilon}^{\mu\nu\rho\sigma} R_{\lambda\nu\rho\sigma} \Pi_{\mu}^{ \ \lambda}+ \bar{\varepsilon}^{\lambda\nu\rho\sigma} R_{\kappa\mu\rho\sigma}  \Pi_\lambda^{ \ \kappa} \Pi_\nu^{ \ \mu} \bigg\}
\\& + 4  \gamma_0^4 \bigg\{\frac{1}{3}\bar{\varepsilon}^{\nu\rho \sigma\alpha} R_{\nu\rho\sigma\alpha} \qty(\langle\Pi^2\rangle - \langle\Pi\rangle [\Pi]) + [\Pi] \langle \bar{\varepsilon}^{\mu\rho \sigma\alpha} R_{\lambda\rho\sigma\alpha} \Pi_{ \mu \kappa}  \rangle
\\& - \langle \bar{\varepsilon}^{\nu\rho \sigma\alpha} R_{\mu\rho\sigma\alpha} \Pi^\mu_{ \ \kappa} \Pi_{ \nu \lambda}  \rangle -  \langle \bar{\varepsilon}^{\nu\rho \sigma\alpha} R_{\lambda\rho\sigma\alpha} \Pi^\mu_{ \ \kappa} \Pi_{ \nu \mu}  \rangle
\\& + \langle\Pi\rangle \bar{\varepsilon}^{\nu\rho \sigma\alpha} R_{\mu\rho\sigma\alpha} \Pi_\nu^{ \ \mu}  -2  \langle \bar{\varepsilon}^{\mu\rho \sigma\alpha} R_{\lambda\nu\sigma\alpha} \Pi_{ \mu \kappa} \Pi^\nu_{ \ \rho}  \rangle \bigg\}.
\label{eq:d18}
\end{split}\end{equation}
From the above it is evident that the disformal transformation contributes extra pieces to the higher-curvature terms which results in altering the effective description. In the case of a general disformal transformation we would have to include several additional contributions. It is evident that these expressions can quickly grow to be very large. Here we focus on the transformation of linearised operators, which are simpler, and leave such considerations for future work.

\section{Next-to-next-to leading order operators} 
\label{appA3}


The usual prescription when working with EFTs is to  write down the most general set of operators consistent with the symmetries of the full theory and use the background equations of motion to eliminate redundant ones as prescribed in \cite{Weinberg:2008hq}. At (NNLO) one may also have to use the first Bianchi identity to relate terms. So far to our knowledge no one has produced the full set of non-redundant six-derivative corrections to gravity, in the context of single field inflation. Here we only tentatively look at some (NNLO) operators. As an example, let us briefly consider the following six-derivative operators

\begin{equation}\begin{split}
S^{(0)}_{NNLO}& =\frac{M^2_{Pl}}{2}  \int \dd^4{x} \sqrt{-  g} \Bigg\{ \frac{b_1}{\Lambda^4}  W_{\mu\nu\rho\sigma}  R^{\nu\sigma}  \phi^{,  \mu}  \phi^{,  \rho} + \frac{b_2}{\Lambda^4}  W_{\mu\nu\rho\sigma}   \phi^{;  \mu  \rho}  \phi^{;  \nu  \sigma} 
\\&+ \epsilon^{\mu\nu\rho\sigma}  \frac{d_1}{\Lambda^4}  W_{\rho\sigma\kappa\lambda}  R^\lambda_{ \ \nu}  \phi^{,  \kappa}  \phi^{,  \mu} \Bigg\}.
\label{eq:AP101}
\end{split}\end{equation}
Following the prescription in Section (\ref{appS21}) and Appendix \ref{appA1} we disformally transform these operators, after we have expanded them to second-order in perturbations of the metric, to find the following contributions

\begin{equation}\begin{split}
S^{(2)}_{NNLO}&=\frac{M^2_{Pl}}{2} \int \dd^4{x} \Bigg\{ \frac{ b_1}{a^2 \Lambda^4} \Bigg[-\frac{(\phi^\prime)^2}{8 c_T^6} \qty(h_{ij}^{\prime\prime})^2 - \frac{\mathcal{H}(\phi^\prime)^2}{4 c_T^6} h_{ij}^\prime h_{ij}^{\prime\prime}  -  \frac{\mathcal{H}(\phi^\prime)^2}{4 c_T^5}  h_{ij}^\prime \nabla^2 h_{ij} 
\\&+ \frac{(\phi^\prime)^2}{8c_T^2} (\nabla^2 h_{ij})^2 \Bigg]
+ \frac{b_2}{a^2 \Lambda^4} \qty[\qty( \frac{\phi^\prime \phi^{\prime\prime}}{4 c_T^6}-\frac{\mathcal{H} (\phi^\prime)^2}{2 c_T^6} ) h_{ij}^\prime h_{ij}^{\prime\prime} + \qty(\frac{\phi^\prime \phi^{\prime\prime}}{4 c_T^5}  - \frac{\mathcal{H} (\phi^\prime)^2}{2 c_T^5} ) h_{ij}^\prime \nabla^2 h_{ij}] 
\\& +   \frac{2 d_1}{a^2 \Lambda^4}  \qty[ \frac{(\phi^\prime)^2 \pd_i h_{lj}^\prime h_{lk}^{\prime\prime}}{ c_T^5} + \frac{2 \mathcal{H}  (\phi^\prime)^2 \pd_i h_{lj}^\prime h_{lk}^\prime}{ c_T^5} - \frac{(\phi^\prime)^2 \pd_i h_{lj}^\prime \nabla^2 h_{lk}}{ c_T^3}  ] \Bigg\}.
\label{eq:AP102}
\end{split}\end{equation} 
The order of the energy expansion is $E^4/\Lambda_*^4$ where, using the arguments in Section (\ref{appS22}), we can deduce that the effective mass scale is given by $\Lambda_*^4= c_T^6 \Lambda^4$. It is easy to see that at the non-relativistic limit $c_T \ll 1$ these operators can pick up enhancements that could stand them relevant to the calculation. Therefore, it may become necessary to consider (NLO) and (NNLO) corrections to gravity. Suppose, we extend the action in (\ref{eq:action}) by including the contributions in (\ref{eq:AP102}). As we've already discussed in Section (\ref{appS33}), when working with a combinations of (NLO) and (NNLO) terms, it is not entirely straightforward to find a field redefinition that can ensure second-order equations of motion. 

Here we take a much more modest approach.
At second-order in perturbations of the metric we expect the disformally transformed action to contain, schematically, the contributions shown in (\ref{eq:1b01}). We recall that this form of the action guarantees second-order equations of motion. 

In the absence of a suitable field redefinition one may produce a theory that leads to at most second-order equations of motion by suitably choosing the free functions characterizing the higher derivative contributions. Of course, as these operators are motivated by a quantum mechanical description of gravity one would expect that such tuning may seem unnatural except if it is protected by some underlying symmetry. While this is true, we believe that there is no great loss of generality in doing things in the way indicated here as long as we maintain the form of the action shown in (\ref{eq:1b01}). Therefore, what follows should be understood as a naive approximation to a much more difficult problem. 

The (NNLO) contributions have been chosen in such way so that the operator coupled to $b_1$ can be used to cancel the contribution $(h_{ij}^{\prime\prime})^2$ in the Weyl squared tensor while the operators coupled to $b_2$ can be used to cancel the contribution $h_{ij}^\prime h_{ij}^{\prime\prime}$ in the operators coupled to $b_1$. Similarly, the operators coupled to $d_1$ can be used to cancel the contribution proportional to $\mathcal{H} h_{ij}^{\prime}$ in the gravitational Chern Simons term and introduce to the action a contribution proportional to $\nabla^2 h_{lk}$. Therefore, with the following definitions

\begin{equation}\begin{split}
b_1=  \frac{4 f_1 \Lambda^2 a^2  c_T^3}{(\phi^\prime)^2}, \quad  b_2 = - \frac{b_1 \mathcal{H} \phi^\prime}{2\mathcal{H} \phi^\prime-\phi^{\prime\prime}} = - \frac{4 f_1 \Lambda^2 \mathcal{H} a^2 c_T^3}{\phi^\prime (2\mathcal{H} \phi^\prime- \phi^{\prime\prime})} \qq{and} d_1= - \frac{2 f_2 \Lambda^2 a^2  c_T^3}{(\phi^\prime)^2},
\label{eq:AP103}
\end{split}\end{equation}
the extended action reduces to 

\begin{equation}\begin{split}
S^{(2)}&=\frac{M^2_{Pl}}{8} \int \dd^3{x} \dd{\eta}  a^2 \Bigg\{  (h_{ij}^\pr)^2 -c_T^2 (\nabla  h_{ij})^2
-\frac{\omega_1}{ a^2  \Lambda^2} (\nabla h_{ij}^\pr)^2 + \frac{\omega_2}{a^2 \Lambda^2}  (\nabla^2 h_{ij})^2
\\& - \epsilon^{ijk}\qty[  \frac{ g_1^\prime}{ a^2\Lambda^2}  (h^q_{ \ i})^\pr(\pd_j h_{kq})^\pr-  \frac{ g_2^\prime}{a^2\Lambda^2}  (\pd^r h^q_{ \ i})\pd_j\pd_r h_{kq}] \Bigg\}.
\label{eq:1b}
\end{split}\end{equation} 
In terms of (\ref{eq:1b01}) we can identify $\alpha=\beta=0, \gamma= \omega_1 (a\Lambda)^{-2},\delta = \omega_2 (a\Lambda)^{-2}, \epsilon = g_1^\prime (a\Lambda)^{-2}, \zeta = g_2^\prime (a\Lambda)^{-2}$, ensuring second-order equations of motion. 

Note, here we do not consider operators of the form $\nabla_\alpha W_{\mu\nu\rho\sigma} \nabla^\alpha W^{\mu\nu\rho\sigma}$ which can have up to three derivatives acting on the metric. Such operators carry contributions to the equations motion with very high momenta which, for the sake of simplicity, we choose to ignore these terms in this work and only partially extend our low-energy action to the higher-derivative regime.

\addcontentsline{toc}{section}{References}

\bibliographystyle{utphys}

\bibliography{bibi} 

\end{document}